\documentclass[usenatbib]{mn2e}   
\pdfoutput=1
\usepackage[pdftex]{graphicx} 
\usepackage{times}
\usepackage{rotate} 
\usepackage{amssymb} 
\usepackage{rotating}
\usepackage{epstopdf} 

\title[The XMM/SDSS  survey]{A Serendipitous  XMM Survey of  the SDSS:
  the evolution of the colour-magnitude diagram of X-ray AGN from $\bf
  z=0.8$  to  $\bf  z=0.1$}  \author[Georgakakis \&  Nandra]  {Antonis
  Georgakakis$^{1}$, K. Nandra$^{2}$ \\ \\ $^1$National Observatory of
  Athens,  V.  Paulou \&  I.  Metaxa,  11532, Greece\\  $^2$Max Planck
  Institut  f\"{u}r Extraterrestrische  Physik,  Giessenbachstra\ss e,
  85748 Garching, Germany\\ }
\begin{document}
\maketitle

\begin{abstract}
A new  serendipitous XMM survey in  the area of the  Sloan Digital Sky
Survey is  described (XMM/SDSS), which  includes features such  as the
merging  of overlapping fields  to increase  the sensitivity  to faint
sources, the use of  a new  parametrisation of  the XMM  point spread
function for the source detection and photometry,
the accurate estimation  of the survey sensitivity.  About 40\,000
X-ray  point  sources   are  detected  over  a  total   area  of  $\rm
122\,deg^2$.   A subsample  of 209  sources detected  in  the 2-8\,keV
spectral  band   with  SDSS  spectroscopic  redshifts   in  the  range
$0.03<z<0.2$,  optical  magnitudes  $r<17.77$\,mag and  $\log  L_X(\rm
2-10\,keV)>41.5$ (erg/s) are selected to explore their distribution on
the   colour   magnitude  diagram.    This   is   compared  with   the
colour-magnitude  diagram   of  X-ray  AGN  in  the   AEGIS  field  at
$z\approx0.8$.    We   find  no   evidence   for   evolution  of   the
rest-frame colours  of X-ray AGN hosts  from $z=0.8$ to  $z=0.1$. This
suggests that the dominant accretion mode of the AGN population, which
is expected to imprint on  the properties of their host galaxies, does
not  change  since  $z=0.8$.   This  argues  against  scenarios  which
attribute the  rapid decline  of the accretion  power of  the Universe
with   time   (1\,dex  since   $z=0.8$)   to   changes   in  the   AGN
fueling/triggering mode.
\end{abstract}

\section{Introduction}\label{sec_intro}

Understanding  the evolution  of Active  Galactic Nuclei  (AGN), which
signpost  accretion  events  onto  Supermassive Black  Holes  (SMBHs),
remains  a challenge for  modern astrophysics.   Although observations
have  demonstrated beyond  any doubt  that the  luminosity  density of
these  systems has  dropped by  more than  1 order  of  magnitude from
$z\approx1$ to the  present day \citep[e.g.][]{Ueda2003, Hasinger2005,
  Ebrero2009, Aird2010}, the physical mechanisms that drive this rapid
decline  are still  not  well  constrained.  The  lack  of a  physical
description for  the cosmological evolution of  SMBHs has implications
that  go  beyond the  AGN  community.   Recent  evidence indicates  an
intimate relation between  the building of galaxies and  the growth of
the SMBH at their centres \citep[e.g.][]{Ferrarese2000, Gebhardt2000}.
Therefore without  a better understanding of AGN  evolution our picture
for the buildup of galaxies will also be incomplete.

Galaxy mergers have long been  proposed as the mechanism that triggers
AGN and drives their  cosmological evolution.  Numerical SPH (Smoothed
Particle  Hydrodynamic)  simulations  demonstrate that  these  violent
events  are very  efficient in  funneling  gas to  the nuclear  galaxy
regions  \citep[e.g.][]{Hernquist1989,Barnes1991,Barnes1996}, where it
can  be   consumed  by  the   SMBH  \citep{Springel2005,DiMatteo2005}.
Consequently in most  semi-analytic cosmological simulations of galaxy
formation  \citep[e.g.][]{Cattaneo2005, Somerville2008,  Wang2008} the
merging of near equal mass  gas rich galaxies is the primary mechanism
for growing SMBHs.   In this family of models the  evolution of AGN is
intimately related  to the  decline of the  fraction of  gaseous major
mergers    with    redshift   \citep[e.g.][]{LopezSanjuan2009}.   
Alternative scenarios propose that the dominant mode of accretion onto
SMBHs changes  with time \citep{Hasinger2008,  Fanidakis2010}, thereby
leading  to  the observed  decline  of  the  AGN space  density  since
$z\approx1-2$. It is suggested for example, that SMBHs form primarily
in   violent   gaseous   major   merger  events   at   high   redshift
\citep[$z\gtrsim1$;   ``QSO-mode''][]{Hopkins2006},  while  stochastic
accretion (e.g.  internal instabilities, minor interactions) dominates
the growth  of SMBHs \citep[``Seyfert-mode''][]{Hopkins_Hernquist2006}
at lower redshift ($z<0.5$) and produces, on average, lower luminosity
systems. Another possibility  proposed by \cite{Fanidakis2010} is that
the dominant AGN fueling mode  shifts from disk instabilities at high
redshift   to  halo   gas  accretion   \citep[also   termed  ``radio''
  mode,][]{Croton2006}   at  low  redshift.    One  of   the  testable
predictions of those models is  that the properties of AGN hosts, such
as  the the  star-formation history  and the  morphology,  change with
redshift,  from $z\gtrsim1$ to  $z<0.5$.  In  contrast, in  the merger
only driven evolution scenario, AGN  should be hosted by galaxies that
have similar properties at all redshifts.

The  study  of  AGN  hosts  as  a  function  of  redshift  requires  a
homogeneously  selected  AGN sample  over  a  wide redshift  baseline,
$z\approx0$  to $z\approx1$  and beyond.   Only then  can  one compare
directly the properties of  AGN hosts at different epochs.  Variations
in the  selection function with redshift  are hard to  account for and
may lead  to erroneous conclusions.   At $z\ga1$ in  particular, X-ray
observations, especially  at energies $>2$\,keV,  are one of  the most
efficient and least  biased methods for locating AGN  with a selection
function  that is  easy  to  quantify.  The  infrared  is a  promising
wavelength regime for finding  active SMBHs, although there are issues
related  to   contamination  of  infrared  selected   AGN  samples  by
starbursts  \citep[e.g.][]{Georgantopoulos2008,  Donley2008, Pope2008,
  Georgakakis2010}. Optical  spectroscopy is also a  powerful tool for
identifying AGN,  but aperture effects, which  are particularly severe
at high redshifts,  raise concerns about dilution of  the AGN emission
lines      by     the      host     galaxy      stellar     population
\citep[e.g.][]{Severgnini2003}.   As a  result, surveys  with  XMM and
Chandra,  both  deep/pencil-beam   and  shallow/wide,  have  been  the
workhorse of the astronomy community for compiling AGN samples.  Those
surveys however, essentially  probe active SMBHs close to  the peak of
the accretion  power of the Universe,  $z\ga 1$, and lack  the area to
provide meaningful  constraints on the  AGN population at  $z\la 0.5$.
The  Sloan  Digital   Sky  Survey  \citep[SDSS;][]{Abazajian2009}  has
identified  the largest  sample  of low  redshift ($z\approx0.1$)  AGN
todate, using  diagnostic emission line  ratios \citep{Kauffmann2004}.
The selection function of that  sample however, is very different from
that of X-ray AGN in deep surveys, rendering the comparison difficult.
For  example, the  SDSS  AGN include  a  large number  of LINERs  (Low
Ionisation  Nuclear  Emission line  Region),  which are  controversial
objects   and  may   not  be   powered  by   accretion  onto   a  SMBH
\citep[e.g.][]{Sarzi2010}.  More  relevant to high  redshift X-ray AGN
surveys are the serendipitous near all-sky AGN samples compiled in the
nearby    Universe   by    the   high    energy    missions   INTEGRAL
\citep[20-100\,keV,  total of  144  Seyfert AGN,][]{Beckmann2009}  and
SWIFT  \citep[15-195\,keV,  total  of  266  Seyferts,][]{Tueller2010}.
Although  those AGN  samples are  selected at  much  higher rest-frame
energies  compared to  X-ray sources  detected  by Chandra  or XMM  at
$z\approx1$ ($<15$\,keV), they are more appropriate than SDSS AGN as a
low redshift  comparison sample.  Ideally  however, one would  like to
select nearby AGN at rest-frame  X-ray energies comparable to those of
Chandra/XMM surveys at $z\approx1$.

In this paper  we describe the compilation of a  large sample of X-ray
selected AGN  at $z\approx0.1$ using a new  serendipitous X-ray survey
(hereafter referred to  as XMM/SDSS) in the area of  the SDSS based on
archival XMM observations. The advantage of the low redshift X-ray AGN
subset of the XMM/SDSS survey is that the selection function is almost
identical to deep pencil-beam samples, thereby minimising differential
selection biases.   As a  result the comparison  between the  AGN host
galaxy  properties between  $z\approx0.1$ and  $z\approx1$  is greatly
facilitated.   The  XMM  archive  includes  over  10  years  worth  of
observations,  which allow  serendipitous  surveys over  many tens  of
square degrees on the sky. Therefore the volume probed by the XMM/SDSS
sample  at  $z\approx0.1$  is  orders  of magnitude  larger  than  any
wide-area  survey  carried  out  by  either XMM  or  Chandra,  thereby
allowing  detailed  studies  of  the  statistical  properties  of  low
redshift  X-ray  AGN.   The  SDSS  is  the field  of  choice  for  our
serendipitous  X-ray survey  as  it provides  the essential  follow-up
observations for  the identification of  low redshift X-ray  AGN.  The
SDSS optical  data include  5-band photometry ($ugriz$)  and extensive
optical spectroscopy for galaxies at $z\approx0.1$.  Wide area surveys
at various wavelengths  in the SDSS are either  underway or completed,
thereby allowing panchromatic studies of the XMM/SDSS X-ray AGN over a
wide redshift baseline. The ancillary data in the SDSS include (i) the
UKIRT Infrared Deep  Sky Survey \citep[UKIDSS,][]{Lawrence2007}, which
will cover $\rm 4000 \, deg^2$ of the SDSS in 4 near-infrared bands to
$K=18.5$\,mag,  (ii)   the  AKARI  all-sky   survey  in  6   mid-  and
far-infrared bands  from $\rm 9-200\mu  m$ \citep{Ishihara2010}, (iii)
the Herschel ATLAS far-infrared  survey \citep{Eales2010} and (iv) the
FIRST  survey, which  provides  deep (1mJy)  radio continuum  (1.4GHz)
images of the entire SDSS \citep{Becker1995}.

The  structure of  the paper  is as  follows. In  sections 2  to  9 we
describe the  automated pipeline developed to reduce  the archival XMM
observations  that overlap with  the SDSS.   The data  reduction steps
include the construction of event files and images in different energy
bands, the  detection of sources,  the calculation and  application of
astrometric   corrections,   the   estimation   of  fluxes   and   the
identification  of  X-ray  sources  with  optical  counterparts.   The
pipeline uses a  new parametrisation of the XMM  Point Spread Function
(Appendix A),  which is employed for source  detection and photometry,
and an  accurate method  for estimating the  sensitivity of  the X-ray
survey to point  sources.  Section 10 demonstrates the  utility of our
serendipitous X-ray AGN  sample by comparing, for the  first time, the
colour magnitude  diagrams of X-ray selected AGN  at $z\approx0.1$ and
$z\approx0.8$.     Section     11    discuss    the     results    and
conclusions. Throughout this paper we adopt  $\rm H_{0} = 100 \, km \,
s^{-1} \, Mpc^{-1}$, $\rm  \Omega_{M} = 0.3$ and $\rm \Omega_{\Lambda}
=   0.7$.    Rest  frame   quantities   (e.g.   absolute   magnitudes,
luminosities) are parametrised by $h=H_{0} / 100$.

\section{Organisation of the XMM data}

\begin{figure}
\begin{center}
\includegraphics[height=0.9\columnwidth]{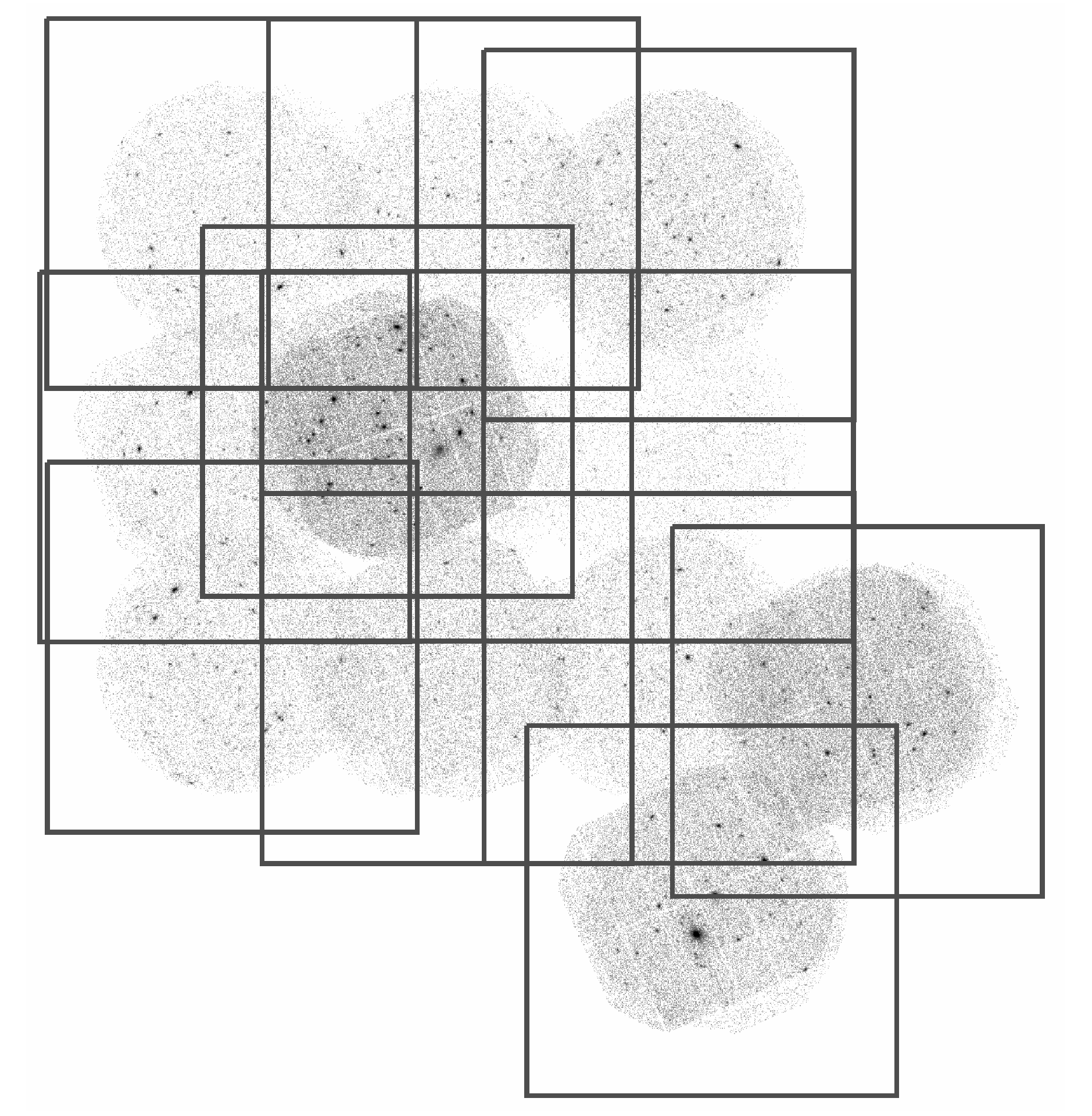}
\end{center}
\caption{Demonstration of how the  XMM observations are organised into
  projects and pointings. The individual XMM observations shown in the
  figure  belong to  the same  project  as they  are contiguous.   The
  identification name of the project is SDSS1341 after the target name
  of the  first osbid that was  assigned to the  project.  The squares
  are  40\,arcmin  on  the  side  and represent  pointings.   All  XMM
  observations that overlap with  a $\rm 40\times40\,arcmin^2$ box are
  merged to  increase the sensitivity  of the serendipitous  survey to
  faint sources.  The positions of  the pointings are defined in a way
  which guarantees that  the entire contiguous area of  the project is
  covered by  at least one pointing  independent of the  layout of the
  individual obsids}\label{fig_pointing}
\end{figure}

For the construction of  the serendipitous X-ray source catalogue, XMM
observations that  were taken  in full frame  and extended  full frame
modes and  were public by July  2009 (total of  3818 observations) are
selected.  We  then chose  only those that  overlap with the  SDSS DR7
\citep[Data  Release  7;][]{Abazajian2009}.    We  use  the  spherical
polygons   \citep{Hamilton_Tegmark2004}  defined   in  the   New  York
University         Value         Added         Galaxy         Catalogue
\citep[NYU-VAGC;][]{Blanton2005} to describe  the geometry of the SDSS
survey.  This  geometry is  the area covered  by the  SDSS-DR7 imaging
survey   after    masking   out   Tycho   stars    as   described   by
\cite{Blanton2005}.   If  the aimpoint  of  an  XMM observation  falls
within one of  the masked areas it is  excluded from further analysis.
This requirement  reduced the number  of analysed XMM  observations to
928.

The  basic unit of  the survey  data structure  is the  individual XMM
observation,  which  will  also  be  referred to  as  ``obsid''.   XMM
observations are assigned to ``projects'' and ``pointings'', which are
defined  below.  Overlapping  obsids  are merged  by  the pipeline  to
maximise the  sensitivity of  the survey to  faint sources.   For that
purpose the  data are organised  into projects which include  at least
one XMM  observation. If  an XMM field  lies within 28\,arcmin  of the
aimpoint of  at least one obsid of  a project, it is  assigned to that
project. Otherwise a new project  is defined.  The radial distance cut
is chosen  to ensure overlap  between adjacent XMM fields,  which have
usable  sizes of approximately  14\,arcmin radius.   Observations that
belong to  separate projects are therefore non  overlapping. A project
is assigned  a unique name using up  to the first 9  characters of the
target  name keyword  of the  first XMM  observation assigned  to that
project. Projects with more than one XMM observations are organised in
pointings, which are defined  as $40\times40 \rm \,arcmin^2$ boxes and
include all the obsids of the  project that overlap with that box. The
first  observation  of a  project  defines  the  centre of  the  first
pointing. Project observations that  lie more than 6\,arcmin away from
an existing pointing define the centres of new pointings.  This choice
of  radial  distance leads  to  substantial  overlap between  adjacent
pointings but ensures that the contiguous area of a project is covered
by  at least  one pointing  independent  of the  layout of  individual
obsids.  This is demonstrated in Figure \ref{fig_pointing} which plots
the  positions of the  pointings of  a particular  project in  the XMM
serendipitous  survey. Once the  pointings of  a project  are defined,
they  are  assigned  all  the  project observations  that  lie  within
28\,arcmin off  the centre  of each pointing.   Therefore all  the XMM
observations that  have at  least a small  fraction of their  field of
view  overlapping with  the $40\times40  \rm \,arcmin^2$  pointing box
belong to the same group and are eventually merged by the pipeline.  A
particular obsid of a project may belong to more than one pointings.

\section{Data processing}


The  XMM observations are  reduced using  the Science  Analysis System
(SAS) version 9.0.  The first step  is to produce event files from the
Observation Data Files (ODF) using the {\sc epchain} and {\sc emchain}
SAS  tasks for  the  EPIC  PN and  MOS  detectors respectively.   This
includes (i)  the creation of raw  event lists and  Good Time Interval
(GTI) data for each CCD of the EPIC detectors, (ii) the identification
of  bad  pixels, (iii)  the  calibration of  the  raw  event lists  by
flagging trailing events, performing pattern recognition, gain and CTI
(Charge  Transfer Inefficiency)  corrections, (iv)  the  estimation of
linearised detector  and sky  coordinates and (v)  the merging  of the
calibrated event  lists of  different CCDs into  event files  for each
EPIC detector.

Pixels along  the edges of  the CCDs of  the PN and MOS  detectors are
removed,  as  in  our  experience  their inclusion  often  results  to
spurious detections.   The excluded pixels for  the EPIC-PN correspond
to rows  with $\rm RAWX =$1,  64 and columns with  $\rm RAWY =199-200$
and  $\rm RAWY <  24$. The  latter set  of pixels  are flagged  out to
reduce bright low energy edges \citep{Watson2009}.  For the EPIC MOS1,
MOS2 we remove pixels with coordinates $\rm (RAWX, RAWY)=(1-4,1-600)$,
$\rm  (RAWX, RAWY)=(597-600,1-600)$,  $\rm  (RAWX, RAWY)=(1-600,1-4)$,
$\rm (RAWX,  RAWY)=(1-600,597-600)$.  Events  with patterns $>  4$ for
the PN  and $> 12$  for the MOS  detectors are removed and  the filter
formulae   (FLAG  \&  0x766b0808)==0   for  the   MOS  and   (FLAG  \&
0x2fb0808)==0 for the PN are  applied. Those flags exclude events from
offset  columns  and  spoiled   frames  and  remove  electronic  noise
\citep{Marty2003}.  Further filtering is applied to observations
targeting bright X-ray sources, such  as QSOs or Galactic stars (total
of 47).   Those observations are  identified by visual  inspection and
are found  to suffer  from diffraction spikes  as well  as out-of-time
events stripes.   The targets  are placed at  the aimpoint of  the XMM
EPIC cameras and therefore the  affected regions are fixed in detector
coordinates.  For  the PN we  remove the out-of-time events  stripe in
rows  $\rm RAWX=29-44$  of CCD4  and the  upper part  of CCDs  4 ($\rm
RAWY>155$)  and  7  ($\rm   RAWY>170$),  which  are  affected  by  the
diffraction spikes  of the  target.  In the  case of MOS1  the removed
pixels  lie on  CCD 1  ($\rm RAWX=292-328$  and  $\rm RAWY=1-600$,$\rm
RAWX=140-480$ and $\rm RAWY=164-454$).  For MOS2 we remove pixels from
CCD  1   with  detector   coordinates  $\rm  RAWX=277-320$   and  $\rm
RAWY=1-600$, $\rm RAWX=163-447$ and $\rm RAWY=144-444$.

\begin{figure}
\begin{center}
\includegraphics[height=0.9\columnwidth]{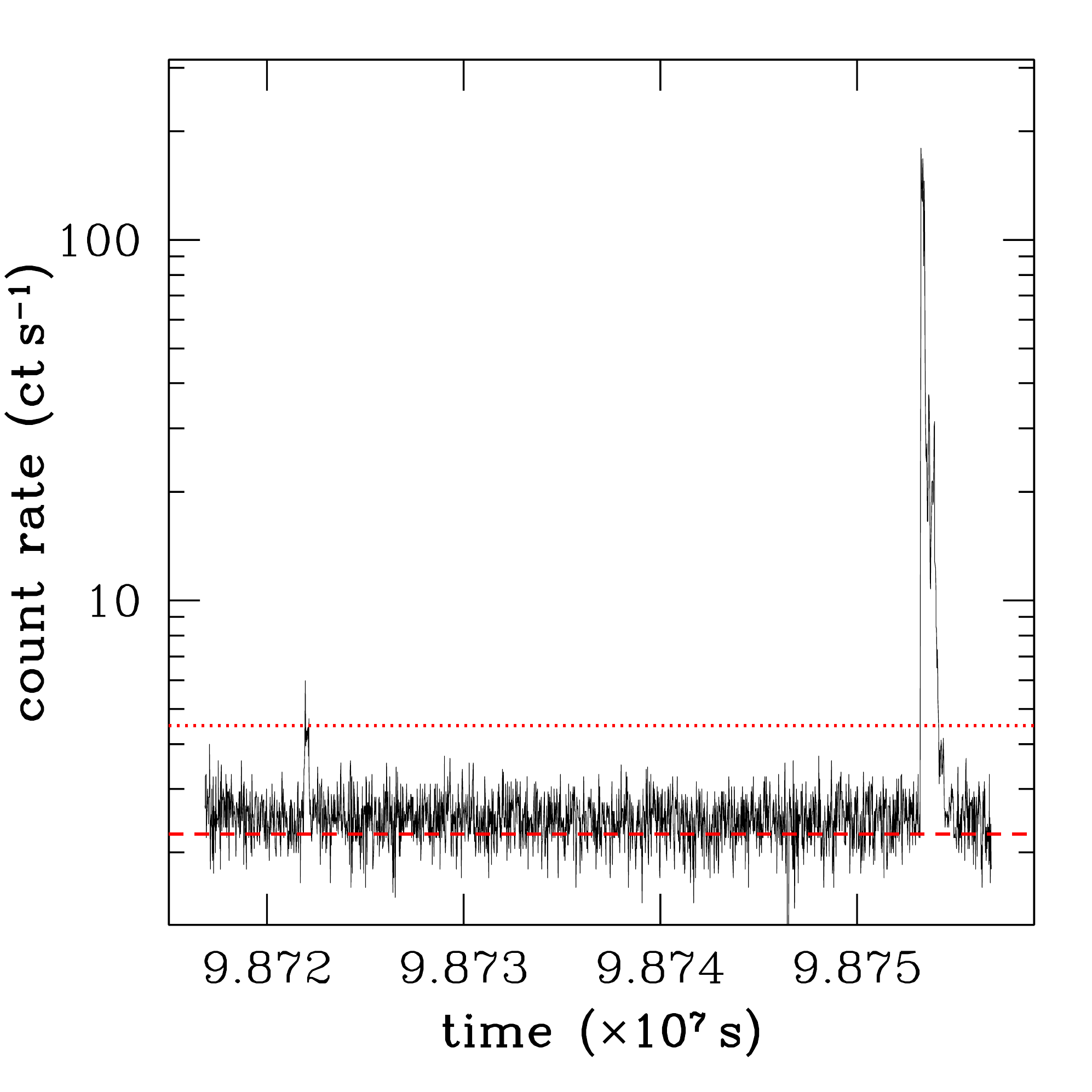}
\end{center}
\caption{Light curve of XMM observation 0041170101 which shows a flare
  at the end of the  observation. The dashed horizontal line marks the
  approximate quiescent background level determined by the methodology
  described in the text. The dotted line corresponds to the count rate
  limit above which events are excluded. }\label{fig_lc}
\end{figure}

Flaring background periods are  identified using a methodology similar
to that described in  \cite{Nandra2007_Fe}. First, sources are removed
from the event file to avoid AGN variability biasing the results.  The
{\sc  ewavelet} task  of SAS  with a  threshold of  5 times  the local
background rms  is used to  detect sources in the  0.5-8\,keV spectral
band  of the  combined  image of  all  EPIC detectors  available to  a
particular XMM  observation.  These sources  are then masked  out from
the event file before generating the background light curve by binning
events  in  the energy  range  0.2-12\,keV  in  20\,s intervals.   The
approximate  quiescent background level  is determined  by calculating
the count  rate limit at which  the excess variance  of the background
light curve is  minimum (c.f.  Nandra et al.   2007, who estimates the
limit  where the excess  variance is  zero).  Periods  with background
count rate 2 times higher than  the level where the excess variance is
minimum  are  excluded  from  the  filtered  event  file  (see  Figure
\ref{fig_lc}).  It is verified  that this  methodology works  well for
XMM.  It fails to identify however, cases where the entire observation
is affected by  flares and hence, has an  overall elevated background.
Therefore  PN and  MOS observations  with quiescent  background levels
higher  than  50  and   $\rm  20\,ct\,s^{-1}$  respectively,  are  not
analysed.   It   is  empirically  found  that   those  values  exclude
problematic  observations from  further analysis.   Also, if  the Good
Time Interval for a particular  detector, after the removal of flaring
periods, is less  than 1\,ks, it is excluded  from further analysis. A
total of 95  obsids have been removed entirely  either because of high
flares or short clean exposure times.

\section{Image construction}\label{sec_image}

Images in  celestial coordinates with  pixel size of  4.35\,arcsec are
constructed in  5 energy bands, 0.5-8\,keV  (full), 0.5-2\,keV (soft),
2-8\,keV  (hard), 5-8\,keV  (very hard;  vhrd) and  7.5-12\,keV (ultra
hard; uhrd).  The upper energy limit  of 8\,keV for the full, hard and
vhrd band  images is chosen because  of the reduced  effective area of
XMM at  higher energies.  Also,  below 0.5\,keV the XMM  background is
elevated    due   to   the    Galactic   X-ray    emission   component
\citep{Lumb2002}.   Therefore,   photons  with  energies   $<0.5$  and
$>8$\,keV  primarily increase the  background and  do not  improve the
signal--to--noise  ratio  of the  final  image,  thereby limiting  the
sensitivity  of   the  survey.   Exposure  maps,   which  account  for
vignetting, CCD gaps and  bad pixels, are constructed at monochromatic
energies of 2\,keV (full band), 1\,keV (soft), 3.5\,keV (hard), 6\,keV
(vhrd)  and 8.5\,keV  (uhrd).   Those values  correspond  to the  mean
photon energy in each spectral band of a power-law X-ray spectrum with
$\Gamma=1.4$  \citep[i.e.  similar  to the  diffuse  X-ray background;
  e.g.][]{Hickox2006} folded  through the sensitivity of  the XMM EPIC
cameras (mirror effective area, energy redistribution matrix).

For projects with at least  one pointing, the EPIC images and exposure
maps of all the observations of  a pointing are merged prior to source
detection to maximise  the sensitivity to faint sources.   In the case
of projects that  consist of a single XMM  observation the EPIC images
and exposure maps are coadded.

\section{Point source detection}

X-ray  point sources are  searched for  in the  merged EPIC  images of
either a  pointing or  a project with  a single XMM  observation.  The
detection  of sources  is performed  independently  in each  of the  5
spectral bands  defined in section \ref{sec_image},  soft, full, hard,
vhrd  and   uhrd.   A  methodology   similar  to  that   described  in
\cite{Nandra2005} and \cite{Laird2009}  is adopted.  Source candidates
are identified using the wavelet based {\sc ewavelet} source detection
task  of  {\sc  sas}  at  a  low  threshold  of  $4\sigma$  above  the
background, where $\sigma$ is the rms of the background counts.

For  each  candidate  source  the  Poisson  probability  of  a  random
background   fluctuation  is  estimated.    This  step   involved  the
extraction of the  total counts at the position of  the source and the
determination of  the local background  value.  For the  extraction of
counts a new set of Point Spread Function (PSF) estimates is employed,
which accounts for  both the varying elliptical shape  and size of the
XMM PSF as  a function of position on the  EPIC detectors.  Details of
the   calculation  of   the  PSF   are  presented   in   the  appendix
\ref{sec_psf}.   At  each source  position  the  counts are  extracted
within an elliptical  aperture that corresponds to 70  per cent of the
PSF Encircled Energy Fraction (EEF).   The total counts at a candidate
source  position,  $T$,  is  the  sum of  the  extracted  counts  from
individual  EPIC cameras.   For each  source the  local  background is
estimated by first masking out  all detections within 4\,arcmin of the
source position  using an elliptical aperture that  corresponds to the
80 per cent EEF ellipse.   The counts from individual EPIC cameras are
then  extracted using  elliptical annuli  centred on  the  source with
inner  and  outer  semi-major  axes  of  5 and  15  pixels  (0.36  and
1.09\,arcmin) respectively.  The position angle and ellipticity of the
ellipses are  set equal  to those of  the 70  per cent EEF.   The mean
local background, $B$, is then  estimated by summing up the background
counts from  individual EPIC  cameras after scaling  them down  to the
area of  the source count extraction region.   The Poisson probability
$P(T,B)$ that the extracted counts  at the source position, $T$, are a
random fluctuation  of the background  is calculated.  We  consider as
sources the  detections with $P(T,B)<4\times10^{-6}$.   At this cutoff
less than about 0.01 false sources per XMM pointing are expected given
the typical size of the 70 per cent EEF aperture (15-20\,arcsec radius
depending on energy  and off-axis angle) and the field  of view of the
XMM (about 14\,arcmin radius).  The candidate source list generated by
{\sc ewavelet} task includes a large number of background fluctuations
which are masked out during the estimation of the local background for
individual sources.  This may lead  to an underestimation of the local
background,  as positive  background fluctuations  may be  excluded as
source candidates.   We account for  this potential source of  bias by
repeating the process  above using as input candidate  source list the
one with Poisson probability $P(T,B)<4\times10^{-6}$.

Our methodology is  optimised for the detection of  point sources. The
final  catalogue however, includes  extended X-ray  sources associated
with hot gas from galaxy  clusters or groups. Also, the extended X-ray
emission  regions of  bright clusters  are often  split  into multiple
spurious detections  by our source detection pipeline.   The {\sc sas}
task  {\sc emldetect}, although  also optimised  for the  detection of
point sources,  can also identify extended X-ray  emission (see Watson
et  al.  2009 for  a discussion  of limitations)  as well  as spurious
detections.  This task is used to perform simultaneous PSF fits to the
source count distribution of all the EPIC images of an XMM observation
to estimate the detection likelihood, the source extent and the extent
likelihood.  {\sc emldetect} is applied separately to each of the five
energy bands,  soft, full, hard,  vhrd and uhrd.  The  free parameters
are the source count rate and source extent.  The source positions are
kept  fixed  during the  minimisation  process.   The background  maps
described in section 7 are used  as input to the {\sc emldetect} task.
The  energy and  position dependent  PSF provided  in  the calibration
database is used in the fit. We caution that the PSFs employed by {\sc
  emldetect}   are  different   from  those   described   in  Appendix
\ref{sec_psf}.   The  task   is  allowed  to  fit  up   to  3  sources
simultaneously if they  fall within the 90 per cent  EEF radius of the
calibration database  PSF.  For the  determination of the extent  of a
source  the PSF  is convolved  with a  $\beta$-model profile.   If the
likelihood  of the  extended model  is below  3 or  the extent  of the
source  is less  than  1.5  pixels then  point  source parameters  are
determined  by the  task.   In  the case  of  pointings with  multiple
observations  the {\sc  emldetect}  task is  ran  on each  observation
separately.  The  individual likelihoods for  a given source  are then
summed   up   and   transformed   to  equivalent   likelihoods   $L_2$
corresponding to the case of  two free parameters, as described in the
{\sc emldetect} documentation.  Based  on visual inspection of the XMM
images we consider as extended those sources with extension likelihood
$>3$ in either the full or the soft spectral bands.  For point sources
we  wish to  keep the  sample selection  as clean  and  transparent as
possible.   One  of  the  advantages  of  our  point-source  detection
methodology  is that  the selection  function, which  is based  on the
Poisson spurious  probability threshold, can  be quantified to  a high
degree of  accuracy (see sensitivity map  section below).  Introducing
additional  selection criteria,  e.g.  keeping  sources above  a given
{\sc  emldetect}  detection  likelihood  threshold, would  modify  the
selection function  of the sample in  a way that is  hard to quantify.
We therefore choose  to keep those biases to  the minimum by excluding
only those point sources for which {\sc emldetect} failed to determine
a reliable  fit and did not  estimate a detection  likelihood.  In the
final  XMM/SDSS  catalogue  those  sources  are listed  with  an  {\sc
  emldetect}  detection likelihood $<0$  and are  considered spurious.
Such sources  are typically associated  with image artifacts,  such as
diffraction spikes  that have not been properly  removed, extended hot
gas emission  split by the  detection algorithm into  multiple sources
etc.

The source catalogues in different  energy bands are merged to produce
a unique list of sources for either a pointing or a project consisting
of a single XMM observation.  A source detected in a particular energy
band is  cross-matched with sources  in other energy bands  by finding
the closest counterpart.  The search radius  is set to the 70 per cent
EEF  circular radius, i.e.   $\rm SMA\times  \sqrt{1-\epsilon}$, where
SMA is  the semi-major axis and  $\epsilon$ is the  ellipticity of the
ellipse.  If a source lies  on multiple obsids or detectors the search
radius  is  estimated  as   the  exposure  time  weighted  average  of
individual 70 per  cent EEF circular radii.  It  is verified that this
choice of search radius is the optimal for identifying the same source
in different energy  bands.  In practice we start  by band-merging the
full band sources and then proceed  with the soft, hard, vhrd and uhrd
catalogues.

As there  is overlap  between adjacent pointings  of a project,  it is
essential to  identify and remove duplicates to  produce unique source
lists  for projects  with multiple  pointings.  For  each source  of a
pointing we  search for the closest counterpart  in adjacent pointings
within  a  fixed  search  radius  of 12\,arcsec.   From  the  list  of
duplicates we keep in the final  list the one with the smaller angular
distance from the aimpoint of the pointing on which it is detected.

In the final band-merged source  catalogue of unique sources there are
1258  extended and 827  spurious sources  and a  total of  39830 point
sources.   Table  \ref{tab_obs} presents  the  total  number of  point
sources detected  in each spectral  band.  It should be  stressed that
our detection method is geared  toward point sources and therefore the
extended source catalogue should be treated with caution.  Providing a
well defined sample of X-ray  selected clusters is {\it not} among our
goals. The final point source catalogue also includes targets.

\begin{figure}
\begin{center}
\includegraphics[height=0.9\columnwidth]{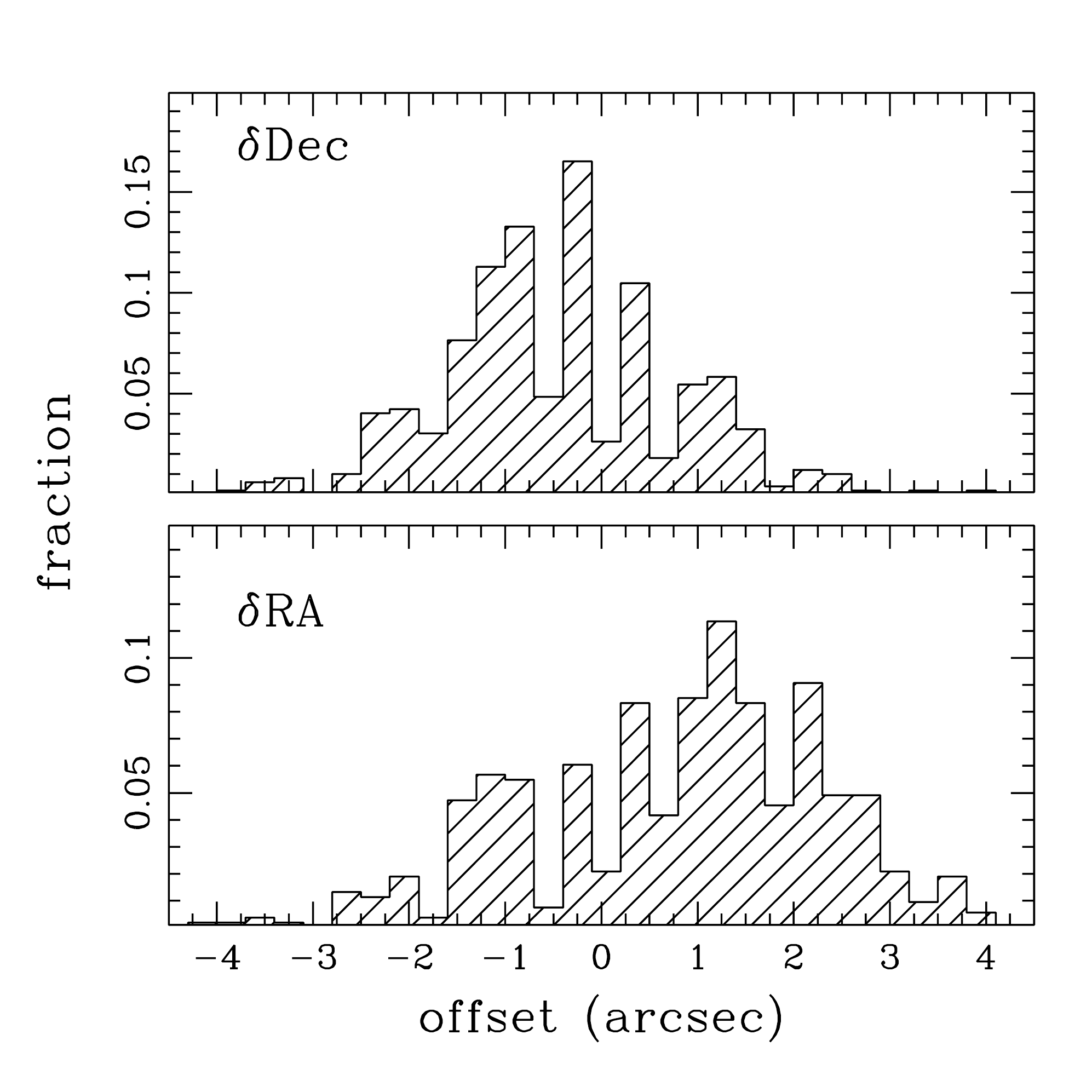}
\end{center}
\caption{Distribution  of the RA  (bottom panel)  and DEC  (top panel)
  corrections  applied  to  the  celestial coordinates  of  the  X-ray
  sources to account for systematic errors in the determination of the
  absolute XMM  pointing.  The  $\delta \rm RA$  and $\delta  \rm DEC$
  offsets  are estimated by  matching the  positions of  X-ray sources
  with optical counterparts in the SDSS.  }\label{fig_wcscorr}
\end{figure}

\section{Astrometry}

The pointing  errors of the XMM  observations are of the  order of few
arcsec.   The positions of  individual X-ray  sources however,  can be
determined to an  accuracy of 1-2\,arcsec (Waston et  al.  2009).  The
astrometry  of  the  X-ray  catalogue  can therefore  be  improved  by
correcting the RA and Dec  of X-ray sources for systematic errors. For
that purpose the positions of  optically selected sources in the SDSS,
which   have    sub-arcsec   astrometric   uncertainties   \citep[$\la
  0.1$\,arcsec,][]{Pier2003},  are  cross-correlated  with  the  X-ray
source list.  The {\sc eposcorr} task  of SAS is used to determine the
offsets in  RA, Dec and roll  angle of an XMM  pointing that maximises
the number  of optical identifications.  Only X-ray  point sources are
used  in  this  exercise,  i.e.   detections flagged  as  extended  or
spurious  by the {\sc  emldetect} task  of SAS  are excluded  from the
cross-matching.  Also, only SDSS sources brighter than $r=22$\,mag are
matched to  the X-ray source  positions.  In the case  of observations
with less  than 8 optical identifications, no  corrections are applied
to  the X-ray  source positions.   Figure \ref{fig_wcscorr}  plots the
distribution  of  the  RA,   Dec  offsets  applied  to  the  celestial
coordinates of  the X-ray sources  and shows that  typical corrections
are about 1-2\,arcsec. In the case of projects with multiple pointings
the  astrometry   of  individual   obsids  should  be   corrected  for
systematics  prior to merging.   We have  verified however,  that this
approach produces  results that  are similar to  the simpler  and less
time-consuming  method of estimating  astrometric corrections  for the
final source  list of individual  pointings of a project,  i.e.  after
merging.

\section{Flux estimation and hardness ratios}

For sources with Poisson detection probability $<4\times10^{-6}$ in at
least one of the five spectral bands fluxes are estimated in all bands
by summing up the counts within  the 80 per cent EEF of the elliptical
PSFs  described   in  Appendix  \ref{sec_psf}.    Compared  to  source
detection, a larger aperture is adopted for the photometry. The choice
of  radius is  to include  as large  a fraction  of source  photons as
possible, while  at the  same time keeping  the background  noise low.
Also,  for larger  EEFs,  the  radius becomes  large  enough that  the
contribution  of photons from  nearby sources  becomes an  issue.  The
local background at the position of a source is estimated as described
in the previous  section. The counts from all EPIC  cameras of all the
XMM  observations  that  a  source  lies  on are  summed  up  for  the
estimation of  the flux  in a particular  energy band.   The different
response  of  the  EPIC  cameras  is accounted  for  by  adopting  the
appropriate count  to flux conversion factors, ECF.   Suppose a source
on  the detector  $i$  of an  XMM  observation.  The  total number  of
counts, $C_i$, at the source position within the 80 per cent EEF are

\begin{equation}\label{eq_expected_counts}
C_{i}=f_X \times ECF_i \times EEF \times t_i + B_i,
\end{equation}

\noindent where $f_X$ is the source flux, $ECF_i$ is the count to flux
conversion  factor for  the  camera $i$,  $EEF=0.8$  is the  encircled
energy fraction, $t_i$ is the exposure time at the source position and
$B_i$ is the  local background value.  Summing up  over $i$, i.e. over
all  observations and all  EPIC cameras  on which  the source  lies on
gives

\begin{equation}\label{eq_sum_flux}
\sum_i  C_{i}=f_X \times  EEF \times  \sum_i \bigl(  ECF_i  \times t_i
\bigr) + \sum_i B_i.
\end{equation}

\noindent The simplest  approach for estimating the source  flux is to
solve the equation above for $f_X$

\begin{equation}
f_X= \frac{1}{EEF} \, \sum_i \bigl(  C_{i} - B_{i} \bigr) \big/ \sum_i
\bigl( ECF_i \times t_i \bigr) .
\end{equation}

\noindent The  equation above however, ignores the  Eddington bias and
the fact  that there is a  continuous probability  distribution of fluxes
which can produce  the observed counts.  We account  for these effects
by adopting the Bayesian  methodology described in \cite{Laird2009} to
calculate the  source fluxes and $1\sigma$ confidence  limits from the
detected counts. In  brief, the probability of the  source having flux
$f_X$  given the  observed  number  of total  counts  $C$ (source  and
background) follows Poisson statistics

\begin{equation}\label{eq_source} 
P(f_X, C) = \frac{T^C \, e^{-T}}{C!} \, \pi(f_X)
\end{equation} 

\noindent where $T$ is the mean expected total counts in the detection
cell. This quantity is determined  as a function of flux from equation
\ref{eq_expected_counts}. The  last term in  equation \ref{eq_source},
$\pi(f_X)$, reflects our prior knowledge of the distribution of source
fluxes, i.e.  the differential X-ray  source number counts,  which are
described by a double power-law \citep[e.g.][]{Georgakakis2008_sense}.
This  term  accounts for  the  Eddington  bias,  in which  statistical
variations of the observed source counts combined with the steep $\log
N - \log S$ of the X-ray population result in brighter measured fluxes
for  the  detected  sources  compared  to their  intrinsic  ones.  The
differential  X-ray  source  counts   are  approximated  by  a  double
power-law  with  faint  and  bright  end slopes  of  --1.5  and  --2.5
respectively and  a break flux  of $\rm 10^{-14}  \, erg \,  s^{-1} \,
cm^{-2}$.   The  adopted  values  are  representative  of  the  double
power-law parameters estimated for X-ray samples selected in different
energy    bands    \citep[e.g.][]{Georgakakis2008_sense}.     Equation
\ref{eq_source}  is  used to  estimate  the  mode  of the  X-ray  flux
distribution, which  can be calculated analytically.  The  68 per cent
confidence limits  on the flux  are estimated by  integrating equation
\ref{eq_source} as described in \cite{Laird2009}.

The energy to flux conversion factor is estimated for each EPIC camera
separately  assuming  a power-law  X-ray  spectrum with  $\Gamma=1.4$,
i.e. similar to the XRB, absorbed by the appropriate Galactic hydrogen
column  density.    The  latter  is   derived  from  the  HI   map  of
\cite{Kalberla2005} using  the right ascension and  declination of the
aimpoint  of  each XMM  observation  and the  {\sc  nh}  task of  {\sc
  ftools}. The  energy to  flux conversion factors  are such  that the
counts from the soft, full,  hard, vhrd and uhrd bands are transformed
to  fluxes in  the 0.5-2,  0.5-10,  2-10, 5-10  and 7.5-12\,keV  bands
respectively.

Hardness ratios (HRs) are  estimated between the soft (0.5-2\,keV) and
the hard (2-8\,keV) band count  rates, S and H respectively, following
the standard definition

\begin{equation}
\rm HR = \frac{H-S}{H+S}.
\end{equation}

\noindent   The   BEHR    \citep[Bayesian   Estimation   of   Hardness
  Ratios,][]{Park2006} code  is used,  which is designed  to determine
HRs in  the Poisson  regime of  low counts and  to compute  the proper
uncertainty  regardless of  whether  the source  is  detected in  both
passbands or not. Hardness ratios were estimated independently for the
PN,  MOS1 and  MOS2 detectors.   In the  case of  multiple overlapping
observations the  summed source and  background counts as well  as the
total exposure time  from each detector are used as  input to the BEHR
code. The  estimated HRs  are scaled to  the on-axis exposure  time by
setting the  BEHR parameters  {\sc softeff} and  {\sc hardeff}  to the
ratios  of the on-axis  exposure time  over the  exposure time  at the
source position in  the soft and hard bands  respectively. In the case
of  overlapping  observations  these  parameters are  defined  by  the
relation

\begin{equation}
eff = \frac{\sum_i C_{i} \, \frac{t_{on,i}}{t_{i}}}{\sum_i C_{i}},
\end{equation}

\noindent where $eff$ represents either the  {\sc softeff} or  the
{\sc hardeff} BEHR parameter,  the  summation  is  over  all XMM
observations,  $i$,  $C_{i}$ 
represents  the counts  (soft or  hard  band) at  the source  position
(source and  background), $t_{on,i}$ is the on-axis  exposure time and
$t_{i}$ is the exposure time at the source position.

\section{Sensitivity maps}

The construction  of the sensitivity  maps follows the  methodology of
\cite{Georgakakis2008_sense},    which   accurately    estimates   the
probability of detecting  a source with a given  X-ray flux accounting
for observational effects, such  as vignetting, flux estimation biases
(e.g. Eddington bias) and the fraction of spurious sources expected in
any source  catalogue. The important  parameters in this  exercise are
the size  and shape of the  detection cell, which are  well defined in
our source  detection method,  and the Poisson  probability threshold,
$P_{thresh}$,  below  which  an  excess  of  counts  is  considered  a
source. By fixing $P_{thresh}$ one  sets the minimum number of photons
in a cell, $L$, for a formal detection.

The first step is the  estimation of the source-free background across
the  EPIC detectors.   The background  map is  estimated  using custom
routines  to first  remove  the  counts in  the  vicinity of  detected
sources using the  80 per cent EEF elliptical  apertures. Pixel values
in the source  regions are replaced by sampling  from the distribution
of pixel values in the local background regions.  These are defined by
elliptical annuli centred on each  source with inner semi-major axis 5
pixels ($\approx21.8$\,arcsec)  larger than the  80 per cent  EEF size
and widths of 15  pixels ($\approx 62.3$\,arcsec).  The resulting maps
are smoothed  by replacing each pixel  value with the  median within a
sliding box of 20 pixels (87\,arcsec) in size. The smoothed background
maps are  then used to  estimate the mean expected  background counts,
within the detection  cell, $B$ (i.e.  the 70  per cent EEF elliptical
apertures).   In the case  of merged  images from  different detectors
and/or  XMM  observations,  $B$  is  the  sum  of  the  mean  expected
background counts within the detection cell of individual images.  The
cumulative  probability  that  the  observed counts  in  a  particular
detection cell will exceed $L$ is

\begin{equation}\label{equ_poisson} 
P_{B}(\ge L) = \gamma(L, B),
\end{equation} 

\noindent 
where  the function  $\gamma(a,x)$  is the  incomplete gamma  function
defined as

\begin{equation} 
\gamma(a, x)  = \frac{1}{\Gamma(a)} \int_{0}^{x} e^{-t}  \, t^{a-1} \,
dt.
\end{equation} 

\noindent Adopting  a detection threshold $P_{thresh}$  one can invert
equation  \ref{equ_poisson}  numerically  to  estimate  the  (integer)
detection  limit  $L$  for   a  cell  with  mean  expected  background
$B$. Repeating this exercise for  different cells across the image one
can  determine   $L$  as   a  function  of   position  (x,y)   on  the
detector. This 2-D  image of $L$ values is  the sensitivity map.  Note
that the sensitivity  map is independent of the  spectral shape of the
source.

A  useful 1-D  representation  of this  image,  with a  wide range  of
applications, is the  total detector area in which  a source with flux
$f_X$  can  be detected.   The  cumulative  distribution  of the  area
plotted  as function  of $f_X$  is  often referred  to as  sensitivity
curve. This  was constructed adopting a Bayesian  approach as follows.
For a source with flux  $f_X$ and a given spectral shape ($\Gamma=1.4$
in this paper) we can determine the probability of detection in a cell
with detection limit $L$ and  mean background $B$.  The total observed
counts in the cell are $C=S+B$,  where $S$ is the mean expected source
contribution.   $C$  is given  by  equation  \ref{eq_sum_flux} in  the
generic case of merged images from different EPIC detectors and/or XMM
observations.

\begin{figure}
\begin{center}
\includegraphics[height=0.9\columnwidth]{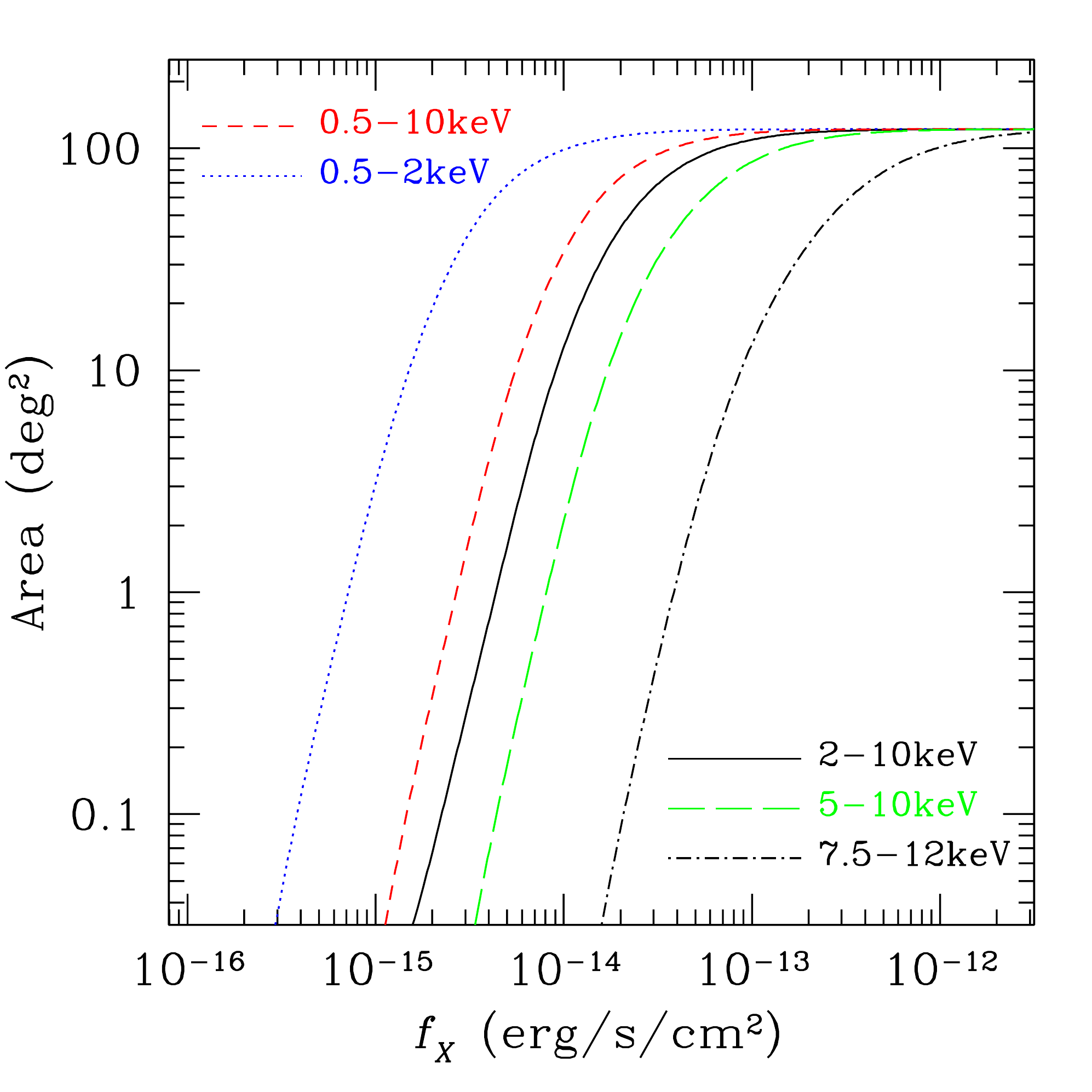}
\end{center}
\caption{Sensitivity  curves  of  the  XMM/SDSS  survey  in  different
  spectral bands.  The curves  are red dashed: full-band; blue dotted:
  soft-band; black continuous: hard-band; green long-dashed: very-hard
  band; black  dashed-dotted: ultra-hard band. The  total surveyed are
  is $\rm 122\,deg^2$.  }\label{fig_curve}
\end{figure}

\begin{figure}
\begin{center}
\includegraphics[height=0.9\columnwidth]{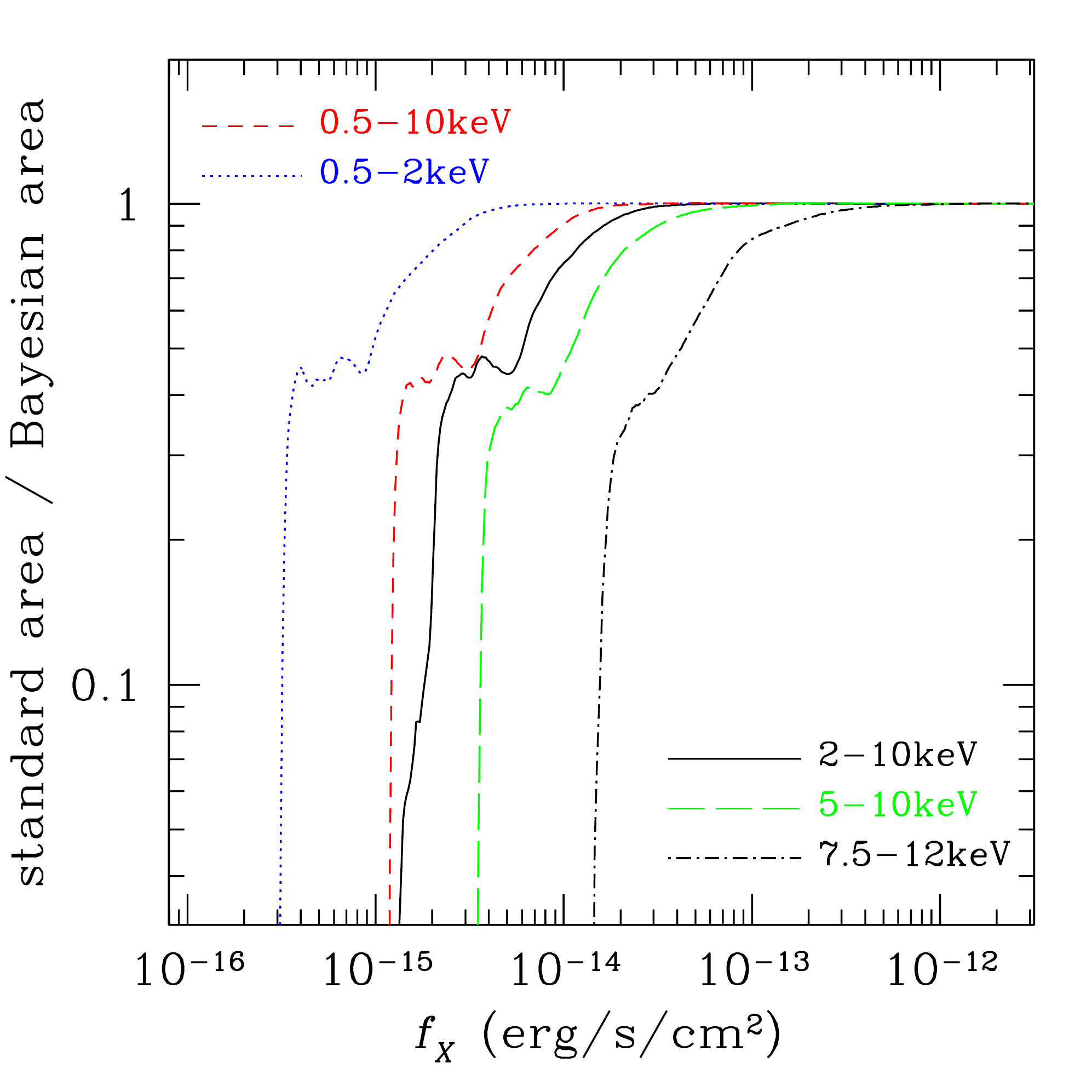}
\end{center}
\caption{Ratio  between the  standard and  Bayesian  sensitivity curve
  estimates as  a function  of flux in  different spectral  bands. The
  curves  are red  dashed:  full-band; blue  dotted: soft-band;  black
  continuous:  hard-band;  green  long-dashed: very-hard  band;  balck
  dashed-dotted: ultra-hard  band.  At  faint fluxes the  standard and
  Bayesian  approaches for  determining the  sensitivity  curve differ
  substantially.}\label{fig_curve_comparison}
\end{figure}

Both   $B$   and   $S$   fluctuate  and   therefore   using   equation
\ref{equ_poisson}  the  probability their  sum  exceeds the  detection
threshold is $P_{C,f_X}(\ge L) = \gamma(L, C)$.  The sensitivity curve
is  the sum  of  the $P_{C,f_X}(\ge  L)$  distributions of  individual
detection  cells. The  sensitivity  curve of  the  XMM/SDSS survey  in
different bands is plotted in Figure \ref{fig_curve}. That figure also
shows that the total XMM/SDSS surveyed area is $\rm 122\,deg^{2}$.  In
addition to the Bayesian  approach we also estimate sensitivity curves
adopting the standard methodology  of assigning a single limiting flux
to a detection cell, i.e.  assuming  that the minimum net counts for a
source   to   be  detected   in   the   cell   are  $S=L-B$.    Figure
\ref{fig_curve_comparison}   compares   the   Bayesian  and   standard
sensitivity  curves in different  spectral bands  and shows  that they
differ substantially at the faint flux end.

\section{Optical Identifications}

The                Likelihood               Ratio               method
\citep[LR;][]{Sutherland_and_Saunders1992,Laird2009}   is  adopted  to
identify X-ray sources with optical counterparts in the SDSS.  In this
exercise the  quantities of interest  are (i) the probability  that an
optical source, at a given distance from the X-ray position and with a
given  optical  magnitude,  is  the  true  counterpart  and  (ii)  the
probability  that  the  same  source  is a  spurious  alignment.   The
likelihood ratio is defined as the ratio between the two probabilities

\begin{equation}\label{eq_lr} {\rm LR}=\frac{q(m)\,f(r)}{n(m)},
\end{equation}

\noindent where  $q(m)$ is the expected magnitude  distribution of the
true  optical  counterparts, $f(r)$  is  the probability  distribution
function of  the positional  uncertainties in both  the X-ray  and the
optical  source catalogues  and $n(m)$  is the  background  density of
optical galaxies of magnitude $m$ in the SDSS $r$-band.

For the positional  accuracy of the X-ray sources  we adopt a Gaussian
distribution with standard deviation  of 1.5\,arcsec (see below).  The
a priori  probability $q(m)$  that an X-ray  source has  a counterpart
with magnitude $m$, is determined  as follows.  Firstly, the X-ray and
optical source  catalogues are matched  by simply finding  the closest
counterpart within a fixed  search radius of 4\,arcsec.  The magnitude
distribution of the spurious matches is estimated by scaling $n(m)$ to
the   area  of   4\,arcsec   radius  within   which   we  search   for
counterparts. This is then  subtracted from the magnitude distribution
of the  counterparts in the  real catalogs to determine  the magnitude
distribution of the true associations, $q(m)$.

Having  estimated   $f(r)$  and   $q(m)$  we  identify   all  possible
counterparts to X-ray sources within  a 4\,arcsec radius and the LR of
each  one is determined  using equation  \ref{eq_lr}.  We  consider as
counterparts those galaxies with LR above a certain limit.  The choice
of  the  cutoff  in LR  is  a  trade  off  between maximum  number  of
counterparts  and minimum  spurious identification  rate. In  order to
assess how  secure an  optical counterpart is  we use  the reliability
parameter defined by \cite{Sutherland_and_Saunders1992}

\begin{equation}\label{eq_rel} \rm Rel_i=\frac{LR_i}{\sum_{j} LR_j
+(1-Q)},
\end{equation}

\noindent  where $\rm  Rel_i$ is  the reliability  of the  $i$ optical
counterpart of  an X-ray source, the  index $j$ of  the summation runs
over all the  possible counterparts within the search  radius and Q is
the fraction  of X-ray sources  with identifications to  the magnitude
limit  of the optical  survey. It  can be  shown that  the sum  of the
reliabilities of individual counterparts equals the expected number of
true associations  (Sutherland \&  Saunders 1992). Comparison  of $\rm
\sum_i  Rel_i$  with  the  total  number  of  counterparts  with  $\rm
LR>LR_{limit}$  provides an  estimate of  the  spurious identification
rate.   By varying  $\rm LR_{limit}$  one can  minimise the  number of
false associations.   We chose to  use $\rm LR_{limit}=1.5$.   At this
cutoff  49 per  cent (19431/39830)  of  the X-ray  point sources  have
counterparts with an estimated spurious identification rate of about 7
per cent.  Table \ref{tab_obs} presents details on the total number of
optical identifications  for different spectral  bands as well  as the
number of sources in  subsmaples defined by applying various selection
criteria,  e.g. magnitude  cuts,  spectroscopic redshift  availability
etc.

\begin{figure}
\begin{center}
\includegraphics[height=0.9\columnwidth]{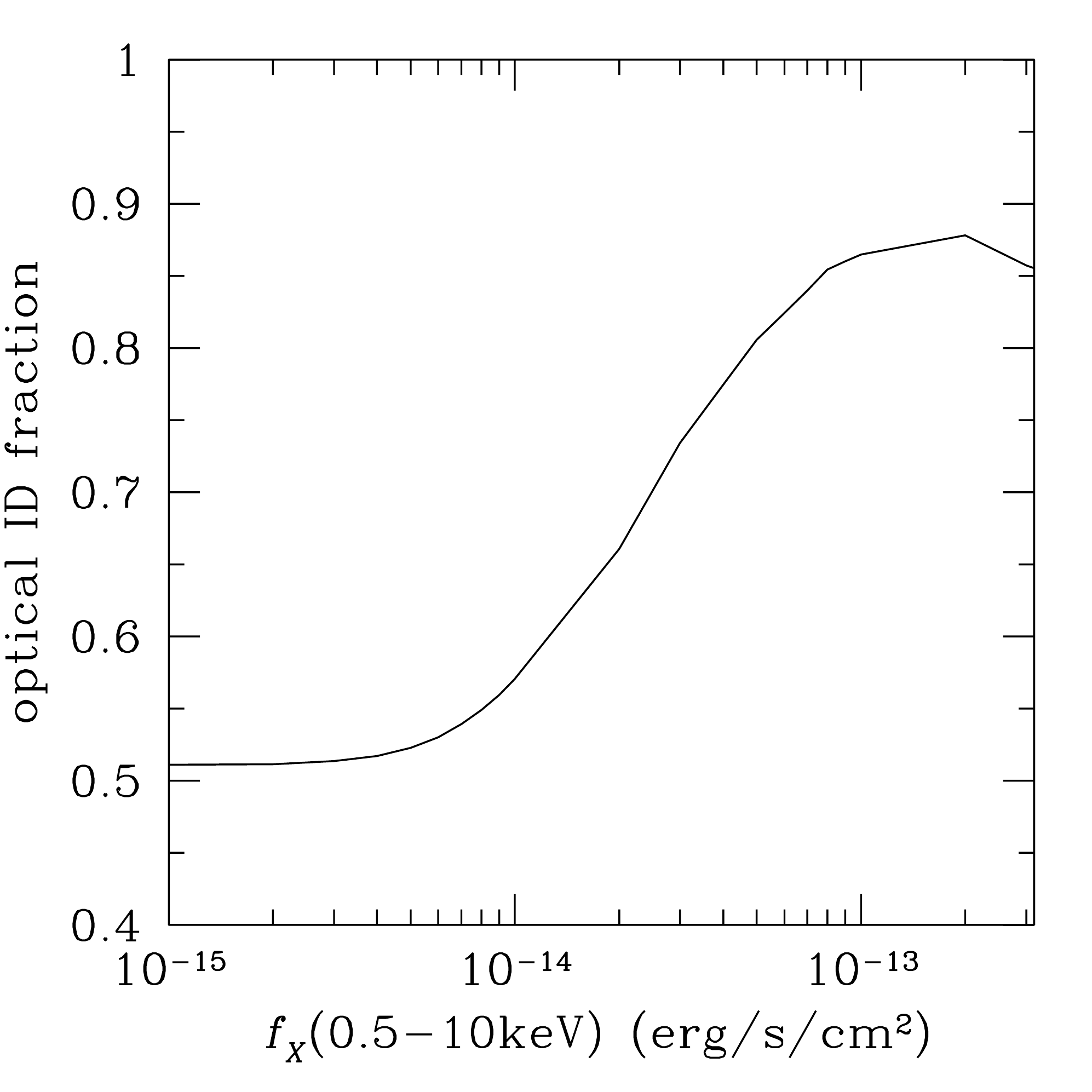}
\end{center}
\caption{Distribution  of  the fraction  of  full-band selected  X-ray
  point sources with optical counterparts as a function of 0.5-10\,keV
  flux limit. }\label{fig_fractionids}
\end{figure}

\begin{figure}
\begin{center}
\includegraphics[height=0.9\columnwidth]{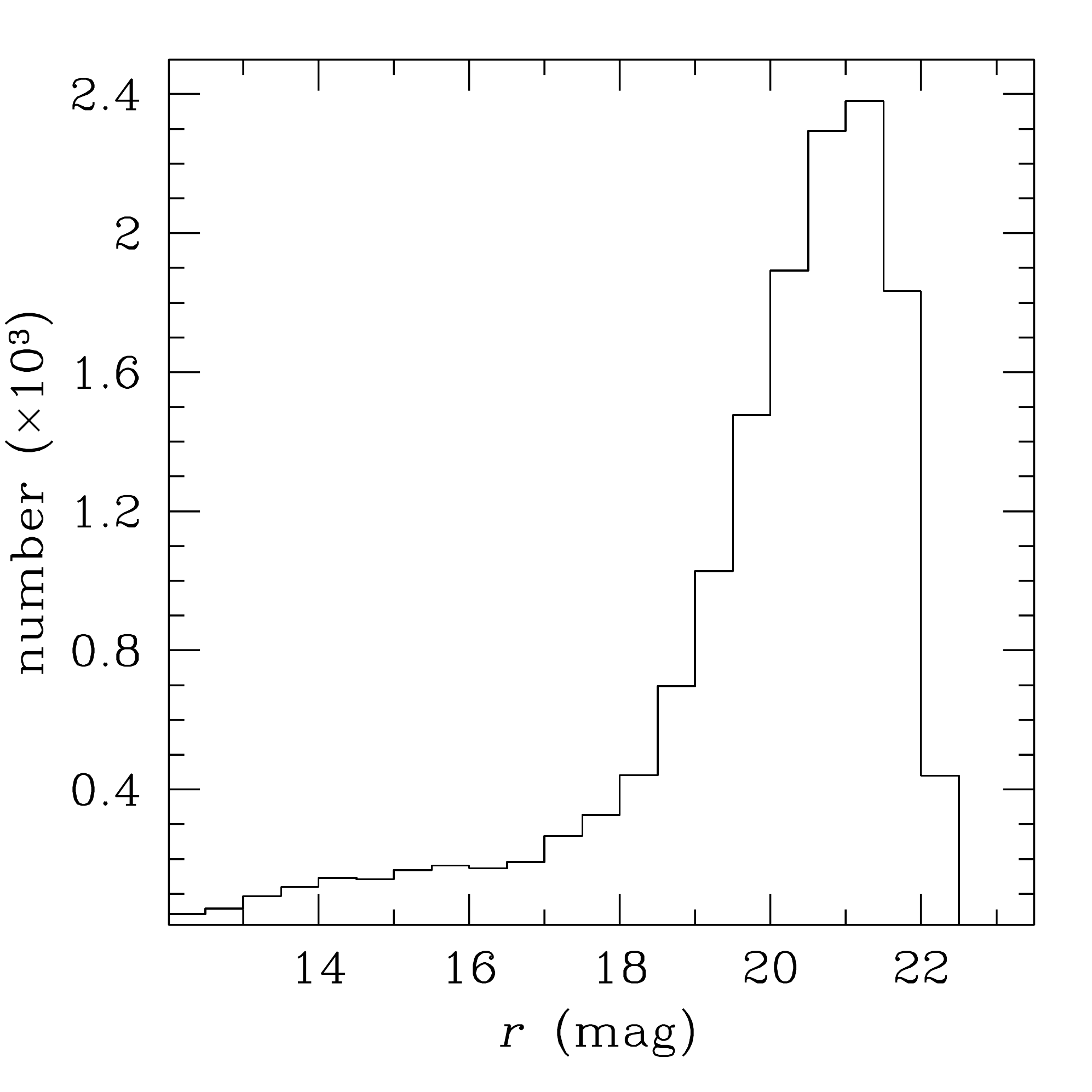}
\end{center}
\caption{SDSS  $r$-band magnitude distribution  of full  band selected
  X-ray sources in the XMM/SDSS survey.}\label{fig_magdist}
\end{figure}

\begin{figure}
\begin{center}
\includegraphics[height=0.9\columnwidth]{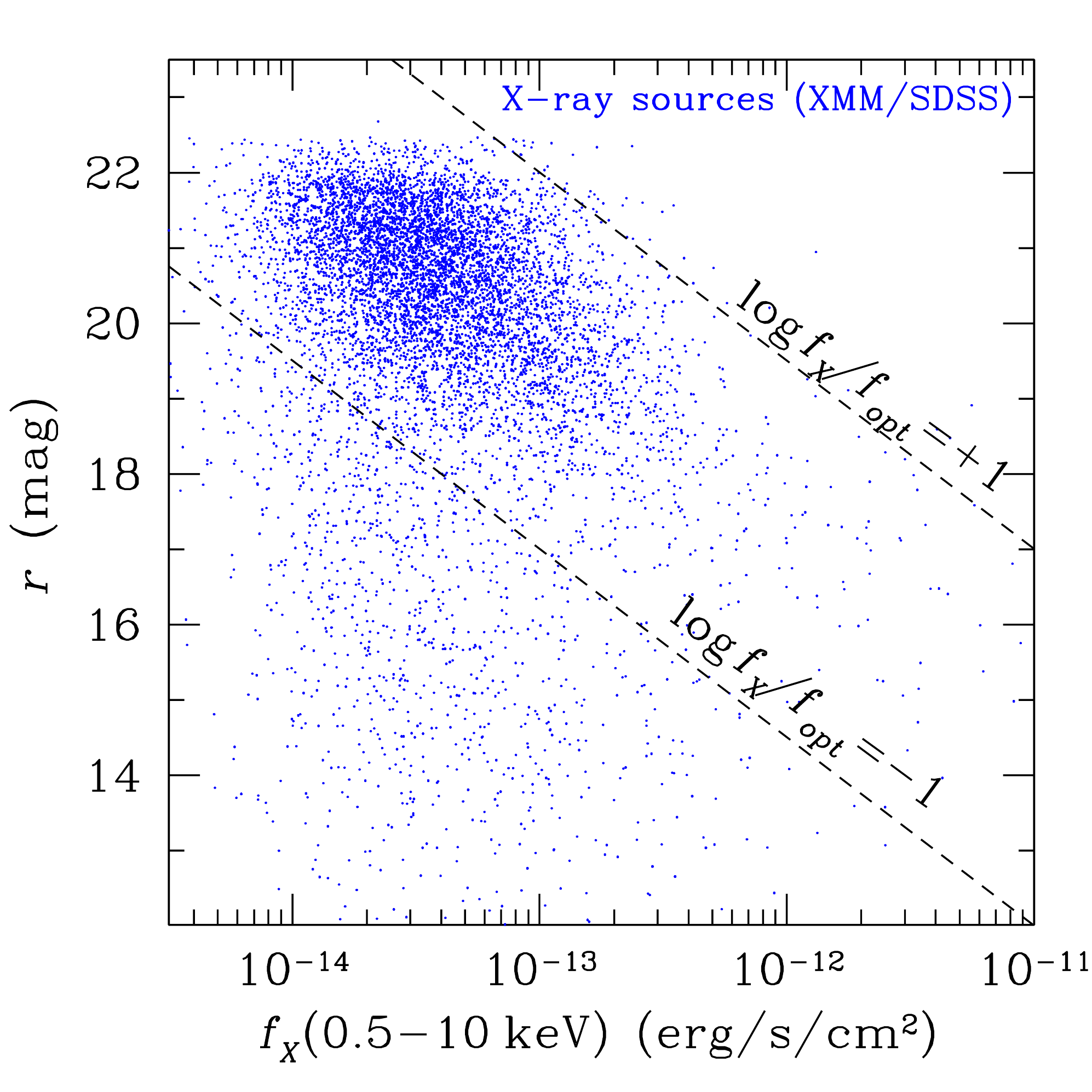}
\end{center}
\caption{$r$-band magnitude against  0.5-10\,keV flux for the XMM/SDSS
  X-ray  sources  with   optical  counterparts.   The  diagonal  lines
  correspond to $\log f_X/f_{opt}=\pm1$.  }\label{fig_fxfopt}
\end{figure}

\begin{figure}
\begin{center}
\includegraphics[height=0.9\columnwidth]{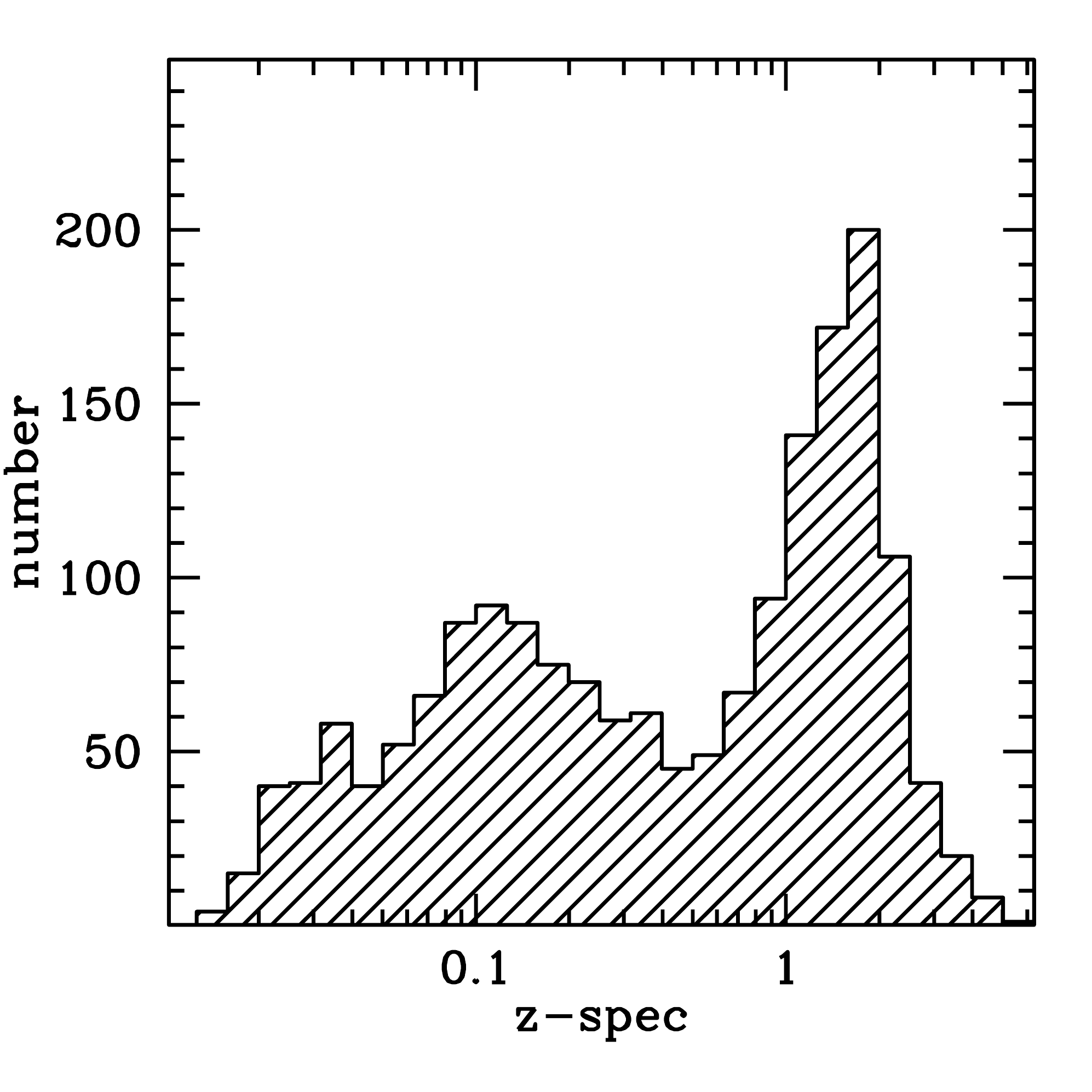}
\end{center}
\caption{Redshift  distribution  of  the  XMM/SDSS  X-ray  sources  in
  logarithmic      bins     of      size     $\rm      \Delta     \log
  z=0.1$.}\label{fig_reddist}
\end{figure}

The optical  identification fraction for  the full band is  plotted as
function  of 0.5-10\,keV flux  limit in  Figure \ref{fig_fractionids}.
The optical identification fraction increases  to about 90 per cent at
bright fluxes, $f_X \approx \rm 10^{-13} \, erg \, s^{-1} \, cm^{-2}$.
The magnitude distribution of the optical counterparts of the XMM/SDSS
sources  is shown  in Figure  \ref{fig_magdist}.  The  distribution of
X-ray sources in the X-ray to optical flux ratio ($f_X/f_{opt}$) plane
is  shown in  Figure \ref{fig_fxfopt}.   A total  of 2067  X-ray point
sources have  spectroscopic observations from the  SDSS.  The redshift
histogram  is shown  in Figure  \ref{fig_reddist}.  For  X-ray sources
associated  with  galaxies (i.e.   optically  resolved) brighter  than
$r<17.77$\,mag   (the  magnitude   cut   of  the   SDSS  main   galaxy
spectroscopic sample) the redshift sampling rate is about 90 per cent,
i.e.   similar to  that of  the  overall SDSS  galaxy population  with
$r<17.77$\,mag.  The majority  of spectroscopically unidentified X-ray
sources  that  are  brighter  than  this  limit  are  associated  with
unresolved optical sources, i.e. they  are most likely distant QSOs or
Galactic stars.

\begin{figure}
\begin{center}
\includegraphics[height=0.9\columnwidth]{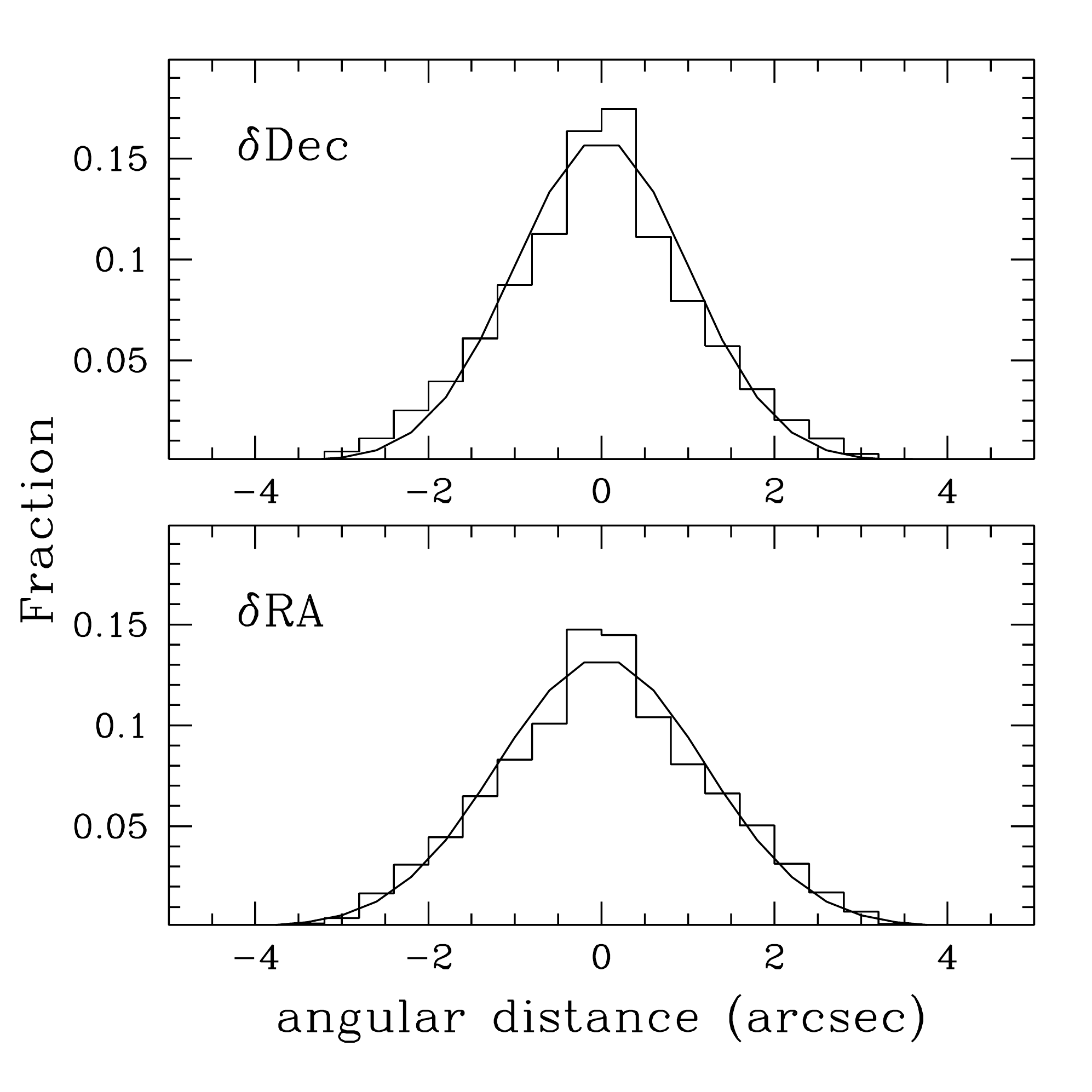}
\end{center}
\caption{Angular separation distribution between the X-ray sources and
  their  optical counterparts  in the  SDSS ($LR>1.5$).   The  top and
  bottom   panels   show   the    Dec   and   RA   offset   histograms
  respectively. The curves are  Gaussians with FWHM of about 1\,arcsec
  and    provide    a    fair    approximation   of    the    observed
  distributions. }\label{fig_dradec}
\end{figure}

Figure \ref{fig_dradec} shows that the distributions of the RA and Dec
angular separations  between the positions of X-ray  sources and their
optical  counterparts ($LR>1.5$)  can be  approximated  with Gaussians
with FWHM of about 1\,arcsec.  The overall astrometric accuracy of the
X-ray catalogue  is therefore  about 1.5\,arcsec.  This  justifies the
use of a normal distribution  with a standard deviation of 1.5\,arcsec
for $f(r)$ in equation \ref{eq_lr}.

The   XMM/SDSS   source   catalogue   is   made   available   to   the
public\footnote{www.astro.noa.gr/$\sim$age}   in  fits   format.   The
information included in that file is listed in Appendix \ref{sec_key}.

\begin{table*}
\caption{X-ray point source catalogue details}\label{tab_obs}
\scriptsize
\begin{tabular}{l c c c c c c}

\hline

Sample &  total number  & $LR>1.5$ &  spectroscopy &  $r<17.77$\,mag &
$r<17.77$\,mag AND & $r<17.77$\,mag AND \\ & of sources & & & & spec-z
& $0.03 \le z_{spec} \le 0.2$ \\ \hline  (1) & (2) & (3) & (4) & (5) &
(6) & (7) \\ \hline

full band & 35727 & 18246 & 1949 & 1920 & 844 & 560 \\

soft band & 33162 & 17631 & 1943 & 2018 & 840 & 552 \\

hard band & 14733 & 8803 & 1237 & 728 & 496 & 298 \\

vhrd band & 2245 & 1722 & 544 & 375 & 297 & 174 \\

uhrd band & 258 & 232 & 143 & 140 & 111 & 58 \\

\hline

\end{tabular} 
\begin{list}{}{}
\item 
The columns are:  (1): X-ray sample; (2): total  number of X-ray point
sources;  (3) sources  with  optical counterpart  ($\rm LR>1.5$);  (4)
X-ray  sources  with $\rm  LR>1.5$  and  SDSS  spectroscopy (5)  X-ray
sources  with   optical  counterparts  brighter   than  $r=17.77$\,mag
(petrosian  magnitude,  AB   system)  after  correcting  for  Galactic
extinction; (6) X-ray  sources with $r\le17.77$\,mag and spectroscopic
redshift  measurement;  (7) X-ray  sources  with $r\le17.77$\,mag  and
redshifts in the range $0.03 - 0.2$, i.e.  the selection criteria used
to construct the colour magnitude  diagram of section 9. Note however,
that in section  9 we use the photometry from  the New York University
Value-Added Galaxy  Catalog (Blanton et al. 2005)  and therefore there
are  small differences  compared in  the sample  size compared  to the
numbers listed in column (7).
\end{list}
\end{table*}

\section{The colour-magnitude diagram}

One of the  main motivations for the XMM/SDSS  serendipitous survey is
the compilation of a low  redshift X-ray selected AGN sample which can
be compared with  high redshift AGN identified in  deep X-ray surveys.
In this  section we present  preliminary results on the  comparison of
the colour-magnitude  diagram (CMD) of X-ray AGN  at $z\approx0.1$ and
$z\approx0.8$.

The  low redshift X-ray  sample consists  of 293  hard-band (2-8\,keV)
detections in the XMM/SDSS survey with $0.03<z<0.2$ and $r<17.77$\,mag
after  correcting for  Galactic extinction  \citep{Schlegel1998}.  The
magnitude cut corresponds to the  limit of the SDSS Main Galaxy Sample
\citep[$r<17.77$\,mag][]{Strauss2002},   which   provides   the   vast
majority of redshifts in the SDSS. The photometry is from the New York
University Value-Added  Galaxy Catalog \citep[NYU-VAGC][]{Blanton2005}
which  corresponds to  the  SDSS DR7  \citep{Abazajian2009}.  This  is
because the  NYU-VAGC provides  better photometric calibration  of the
SDSS data \citep{Padmanabhan2008} compared  to the DR7 and an accurate
description of the  SDSS window function.  A small  number of XMM/SDSS
sources with $0.03<z<0.2$ and $r<17.77$\,mag (total of 5, see column 7
of Table  \ref{tab_obs}) do not  have photometric measurements  in the
NYU-VAGC  and  have  been  excluded  from the  final  sample.   Visual
inspection shows  that all of them are  close to the edge  of the SDSS
field of view.  The choice of the 2-8\,keV X-ray band is because it is
less affected  by obscuration biases  compared to softer  energies and
the XMM's sensitivity remains high in that band thereby resulting in a
sample that is sufficiently large for statistical studies.  The CMD of
the overall galaxy population at $z\approx0.1$ is defined by selecting
a total  of 13,619 NYU-VAGC  sources with $r<17.77$\,mag  that overlap
with the  XMM/SDSS survey fields  and have spectroscopic  redshifts in
the interval $0.03<z<0.2$.

The    {\sc   kcorrect}    version   4.2    routines    developed   by
\cite{Blanton_Roweis2007} are  used to  fit spectral models  to the
NYU-VAGC  $ugriz$ photometry and  then estimate  rest-frame magnitudes
(AB system) in the $^{0.1}u$, $^{0.1}g$ bands, which are the SDSS $u$,
$g$  filters shifted  to $z=0.1$.   The  advantage of  this choice  of
bandpasses  is  twofold.   They   are  close  to  rest-frame  for  our
$z\approx0.1$ XMM/SDSS sample and also have effective wavelengths that
are similar to  the rest-frame effective wavelengths of  the DEEP2 $R$
and $I$ filters at $z=1$ \citep{Blanton2006}.

The  high  redshift CMD  is  constructed  using  data from  the  AEGIS
\citep{Davis2007}. The spectroscopic  sample consists of 6797 galaxies
observed  as part of  the DEEP2  survey (Data  Release 3)  with secure
redshift determinations \citep[$>90$\%  confidence level, quality flag
  $Q  \ge 3$;][]{Davis2007}  in the  interval $0.6<z<1.2$.   The X-ray
data in  the AEGIS are from  the Chandra survey of  the Extended Groth
Strip (AEGIS-X  survey, Laird  et al.  2009).   The X-ray  sources are
identified with DEEP2 galaxies  using the Likelihood Ratio methodology
described in Laird et al.   (2009). The spectroscopy for X-ray sources
is from a variety of sources:  the DEEP2 redshift survey, the SDSS and
a number of spectroscopic programs  that have targeted galaxies in the
original  Groth Strip \citep{Weiner2005}.   We select  a total  of 115
X-ray  sources  in  the  spectroscopic redshift  interval  $0.6<z<1.2$
detected in  the 0.5-7\,keV  band of the  AEGIS-X survey.   The median
redshift of  the sample is 0.8. The  choice of the X-ray  band for the
detection of  sources is  to ensure that  our $z\approx0.8$  sample is
selected   at   similar  rest-frame   energies   (1-14\,keV)  as   the
$z\approx0.1$ sample (2-8\,keV),  thereby facilitating the comparison.
For  the  estimation  of   rest-frame  quantities  we  use  the  $BRI$
photometry obtained as part of the DEEP2 survey \citep{Coil2004}.  The
{\sc kcorrect} routines  are used to fit models to  the $BRI$ data and
then estimate  the rest-frame $^{0.1}u$  and $^{0.1}g$-band magnitudes
from the observed DEEP2 $R$ and $I$-band photometry respectively.

\begin{figure*}
\begin{center}
\includegraphics[height=0.9\columnwidth]{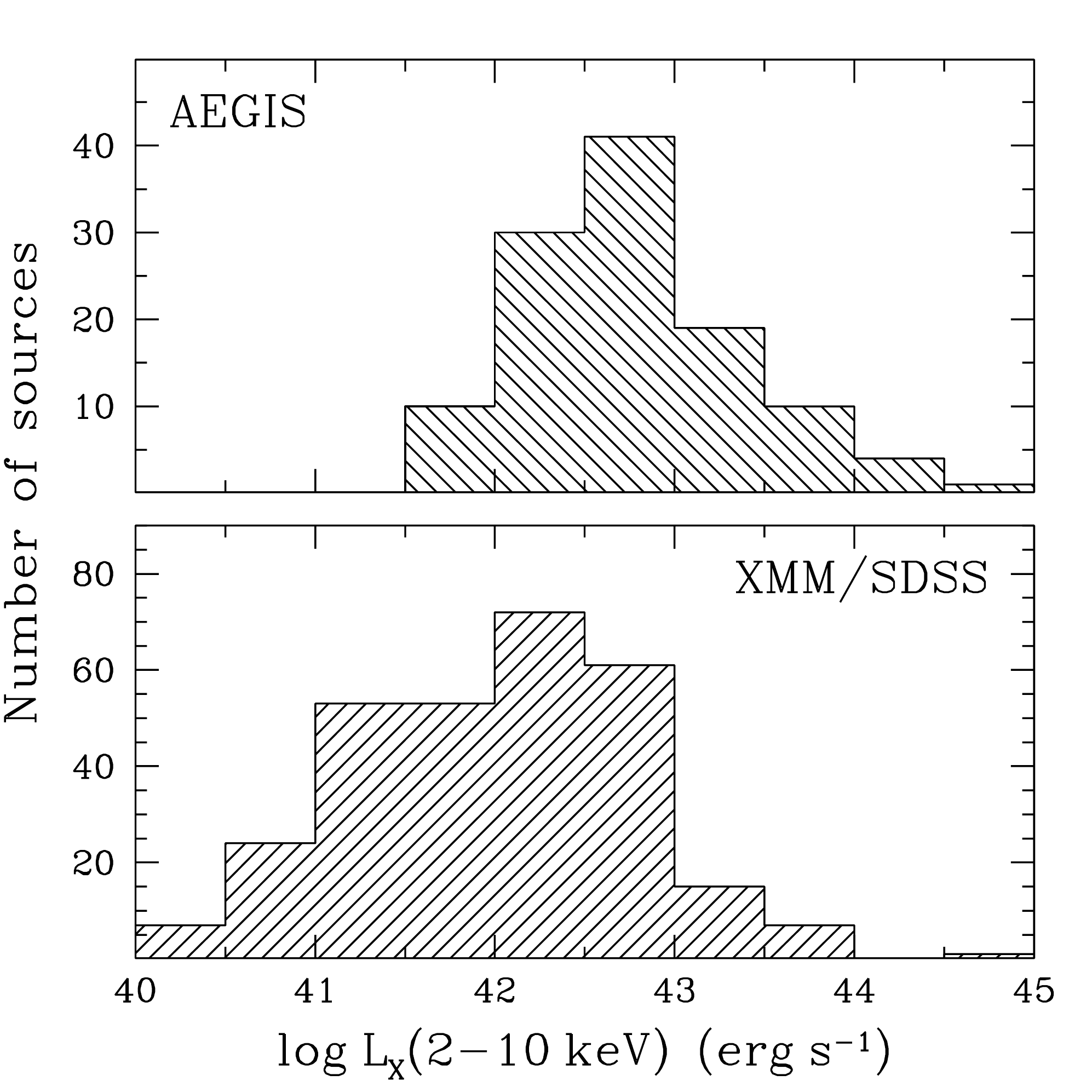}
\includegraphics[height=0.9\columnwidth]{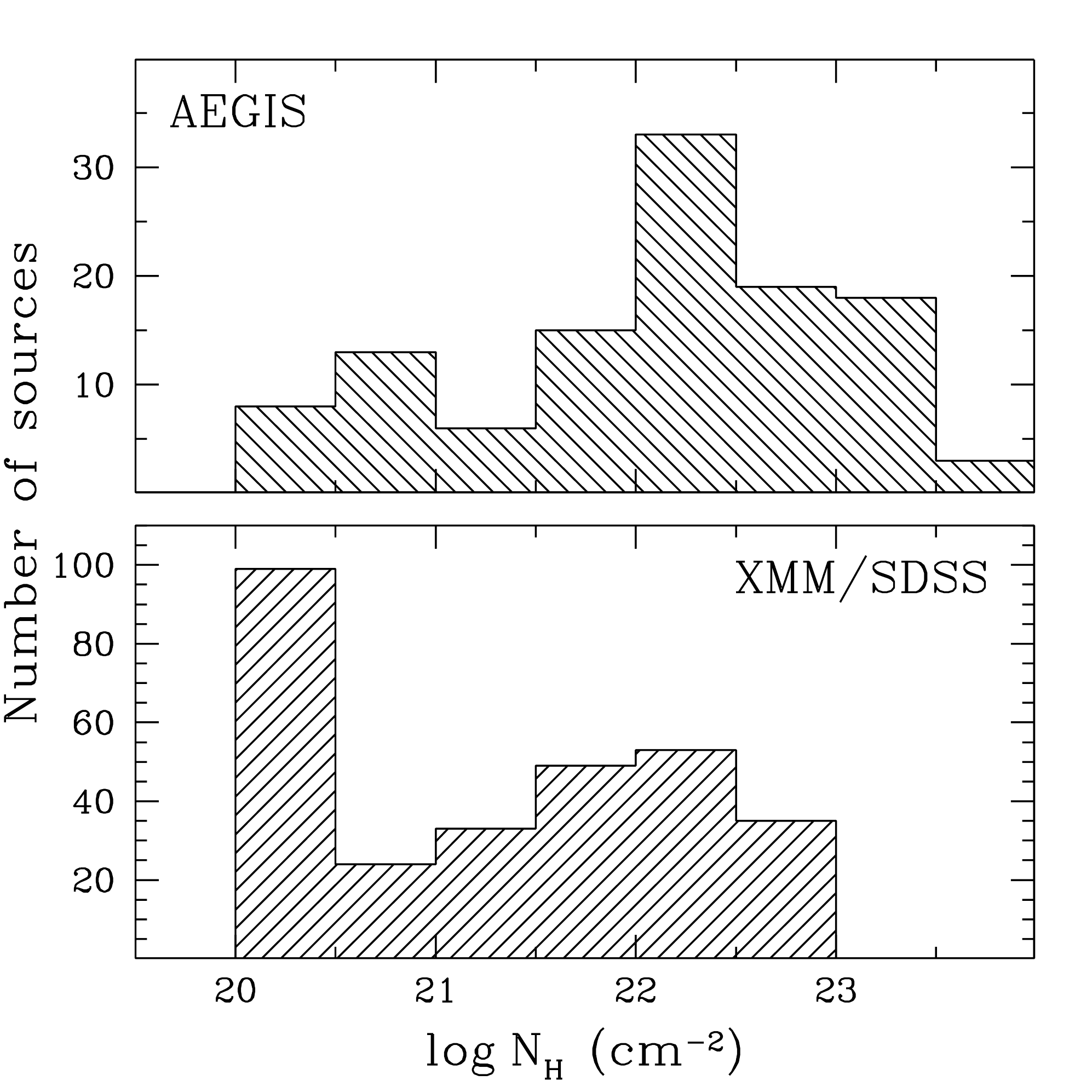}
\end{center}
\caption{{\bf Left panel:}  2-10\,keV X-ray luminosity distribution of
  X-ray sources in the AEGIS (top)  and the low redshift subset of the
  XMM/SDSS (bottom) samples.  The X-ray luminosities are corrected for
  photoelectric  absorption.  In  the  case  of  AEGIS  the  2-10\,keV
  luminosity  is estimated  from the  observed 0.5-7\,keV  count rate.
  The XMM/SDSS sample at $0.03<z<0.2$ reaches luminosities as faint as
  $10^{40} \, erg  \, s^{-1}$, while AEGIS at  $0.6<z<1.2$ probes more
  luminous  X-ray  sources.   {\bf  Right panel:}  Histograms  of  the
  absorbing  hydrogen column  density, $N_H$  in the  AEGIS  (top) and
  $z\approx0.1$  XMM/SDSS (bottom) samples.   This is  determined from
  the observed  hardness ratio of  individual sources.  For  the AEGIS
  the hardness ratio is estimated  from the count rates in the Chandra
  0.5-2  and 2-7\,keV  spectral bands.   In the  case of  the XMM/SDSS
  sample the  hardness ratio  is defined between  the XMM's  0.5-2 and
  2-8\,keV bands.}\label{fig_hist}
\end{figure*}

Figure  \ref{fig_hist}  compares  the absorption  corrected  2-10\,keV
X-ray luminosity and the hydrogen column density ($N_H$) distributions
of X-ray  sources in the AEGIS  sample and the low  redshift subset of
the XMM/SDSS  survey.  The $N_H$ of individual  sources are determined
from the hardness  ratios between the soft (0.5-2\,keV  for both AEGIS
and  XMM/SDSS) and  the hard  (2-7\,keV for  AEGIS, 2-8\,keV  for XMM)
X-ray bands assuming an  intrinsic power-law X-ray spectrum with index
$\Gamma=1.9$ \citep[e.g.][]{Nandra1994}.  The derived column densities
are  then used  to correct  the X-ray  luminosities  for photoelectric
absorption.   The XMM/SDSS  sample  at $z\approx0.1$  is sensitive  to
sources with $L_X  (\rm 2 - 10 \,  keV ) > 10^{40} \,  erg \, s^{-1}$.
In contrast  the AEGIS detects  X-ray sources brighter than  $\log L_X
(\rm 2 - 10  \, keV ) > 41.5$ ($\rm erg  \, s^{-1}$) at $z\approx0.8$.
To avoid luminosity dependent biases  in the comparison of the CMDs of
X-ray sources in the AEGIS and the low redshift subset of the XMM/SDSS
we limit both samples to $\log L_X (\rm 2 - 10 \ , keV ) > 41.5$ ($\rm
erg \, s^{-1}$).  X-ray sources  with luminosities $\rm \la 10^{42} \,
erg  \, s^{-1}$  are often  excluded from  AGN studies  because normal
galaxies are believed to be an important source of contamination below
this limit.   Although normal galaxy  {\em candidates} are  present at
faint  X-ray luminosities,  AGN  remain the  dominant population.   In
\cite{Georgakakis2007}  for example,  X-ray detected  starburst galaxy
candidates represent only  about 20 per cent of  the GOODS-North X-ray
source population in the  luminosity interval $\rm 10^{40} - 10^{42}\,
erg \, s^{-1}$.   This fraction is also likely to be  a upper limit as
some of the  galaxy candidates may turn out to  be low luminosity AGN.
Furthermore,   \cite{Georgakakis2008AN}  cross-correlated   the  2XMMp
catalogue \citep{Watson2009}  with the SDSS  spectroscopic database to
select normal  galaxies at $z\approx0.1$.  Such sources  were found to
represent 20 per cent of the overall X-ray population for luminosities
$\rm 10^{40} - 10^{42}\, erg \, s^{-1}$.  This fraction is also likely
to  be   a  strong  function  of  luminosity   within  that  interval.
Contamination by non-AGN  is therefore not a concern  for the XMM/SDSS
and the AEGIS X-ray source samples with  $\log L_X (\rm 2 - 10 \ , keV
) > 41.5$ ($\rm erg \, s^{-1}$).  This cut reduces the total number of
low  redshift ($0.03<z<0.2$,  $r<17.77$\,mag)  XMM/SDSS X-ray  sources
from 293 to 209.  The sample  includes a total of 16 XMM targets, i.e.
non serendipitous X-ray sources, most  of which (13/16) are type-1 AGN
with  broad optical  emission lines.  We choose  to keep  them  in the
sample as they are not affecting the main results and conclusions.

\begin{figure*}
\begin{center}
\includegraphics[height=0.9\columnwidth]{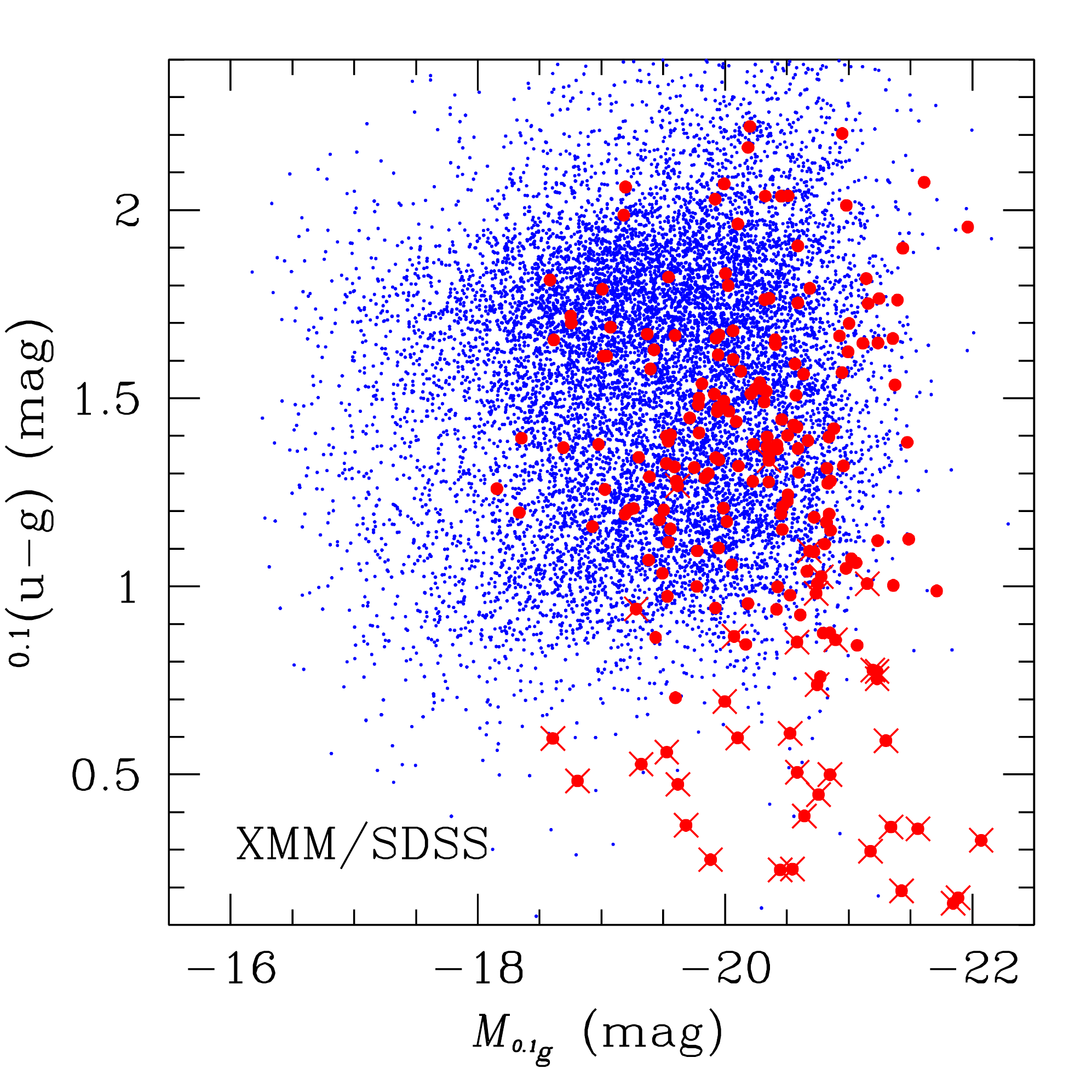}
\includegraphics[height=0.9\columnwidth]{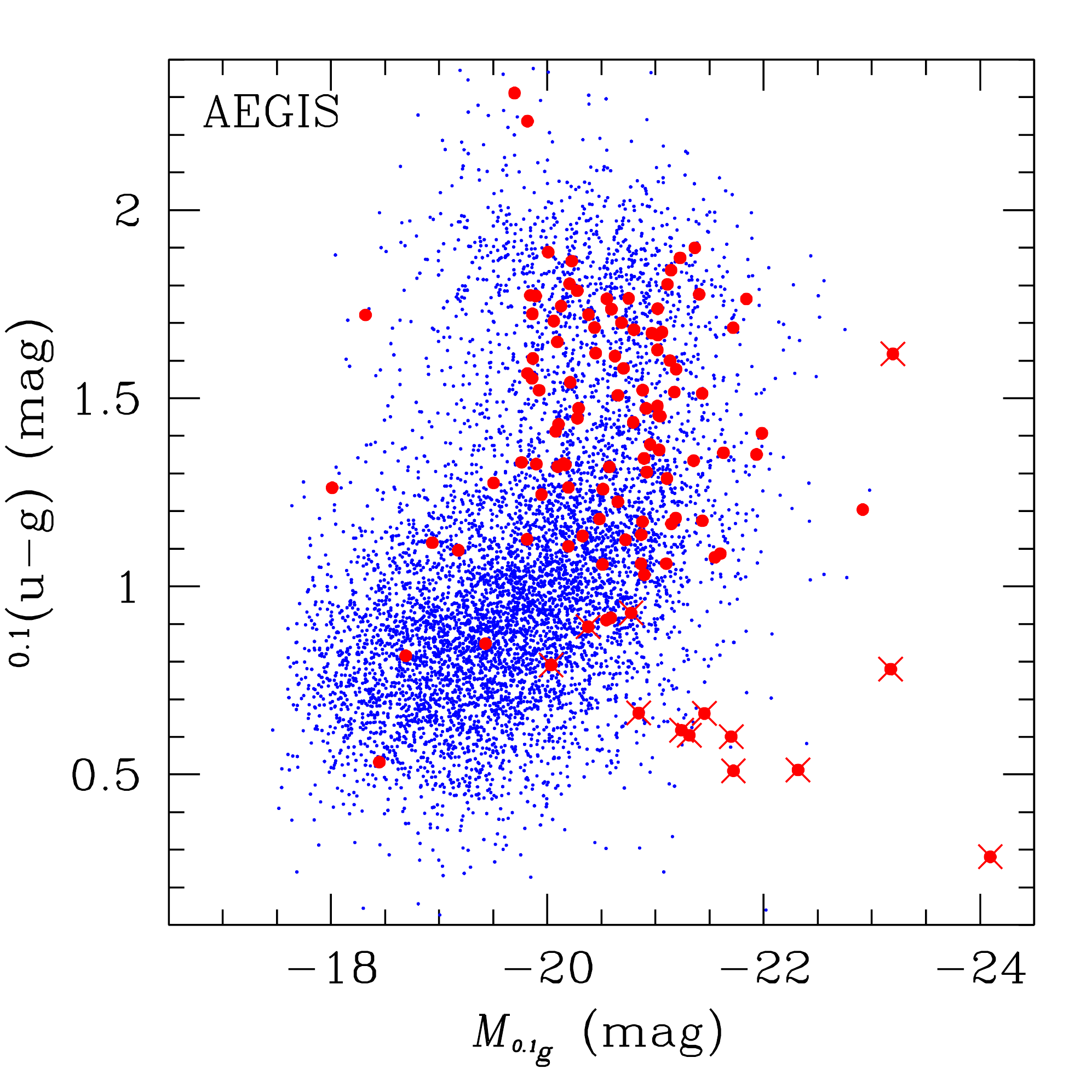}
\end{center}
\caption{$^{0.1}(u-g)$  vs $M_{^{0.1}g}$  Colour magnitude  diagram of
  galaxies (blue dots) and X-ray AGN (red circles) in the XMM/SDSS low
  redshift   sample   ($0.03<z<0.2$;  left   panel)   and  the   AEGIS
  ($0.6<z<1.3$;  right  panel). Crosses  indicate  X-ray sources  with
  rest-frame colours contaminated by AGN light.}\label{fig_cmd}
\end{figure*}

\begin{figure}
\begin{center}
\includegraphics[height=0.9\columnwidth]{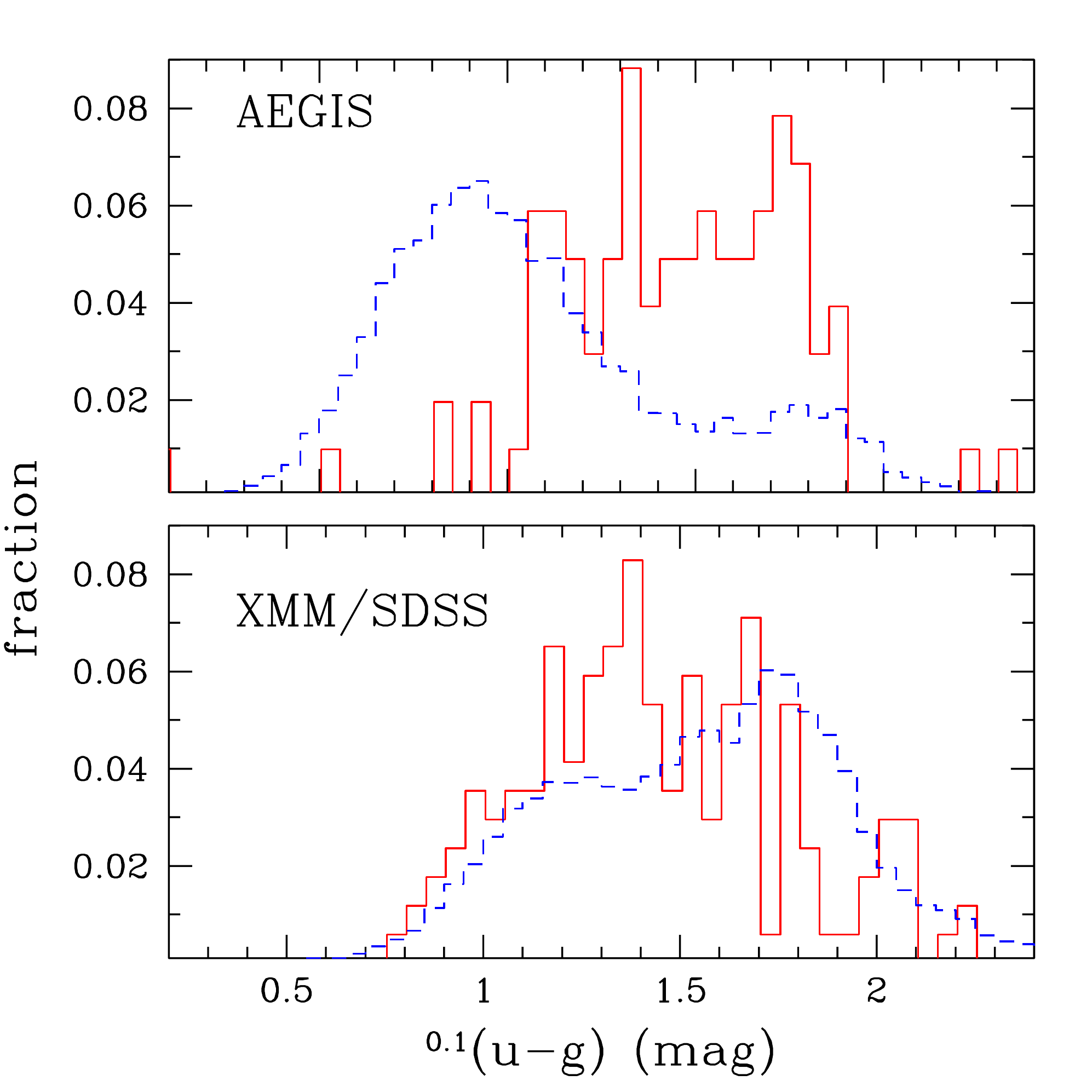}
\end{center}
\caption{$^{0.1}(u-g)$ colour  distribution of galaxies  (blue hatched
  histogram) and X-ray  AGN (red solid histogram) in  the XMM/SDSS low
  redshift   sample  ($0.03<z<0.2$;   lower  panel)   and   the  AEGIS
  ($0.6<z<1.3$; upper panel). All histograms are normalised by the
  total number of sources in each sample.
}\label{fig_hist_cmd}
\end{figure}

Figure \ref{fig_cmd}  compares the $^{0.1}(u-g)$  vs $M_{^{0.1}g}$ CMD
of  galaxies and  X-ray  sources in  the  AEGIS and  the low  redshift
XMM/SDSS  survey  sample.   X-ray  sources  which  either  show  broad
emission  lines in their  optical spectra  and/or a  prominent nuclear
point source in their broadband optical images (SDSS or HST/ACS survey
of  the  AEGIS)  are  marked  in Figure  \ref{fig_cmd}.   The  optical
continuum of those sources is significantly contaminated by AGN light.
Their colours  do not provide  information on their host  galaxies and
are therefore excluded from  further analysis. As expected the optical
colours   of  these   sources  are   very  blue.    The  $^{0.1}(u-g)$
distribution   of    the   two    samples   are   shown    in   Figure
\ref{fig_hist_cmd},  which better demonstrates  the bimodality  of the
galaxy colours in both the AEGIS  and the SDSS.  We set the separation
between red sequence and  blue cloud galaxies at $^{0.1}(u-g)=1.4$ for
both  the  AEGIS  at  $z\approx0.8$  and the  SDSS  at  $z\approx0.1$.
Although the dividing line between  red and blue galaxies also depends
on absolute  magnitude \citep[e.g.][]{Bell2004, Willmer2006},  for our
purposes it is sufficient to  apply a simple colour cut (e.g.  Blanton
2007).  X-ray sources  with colours contaminated by AGN  light are not
plotted in Figure \ref{fig_hist_cmd}. It is interesting that X-ray AGN
in both the  AEGIS and the low redshift subset  of the SDSS/XMM survey
have similar rest-frame  colour distributions.  The Kolmogorov-Smirnov
test estimates a  probability of 50 per cent that  the two samples are
drawn from  the same parent  population.  Also, the fraction  of X-ray
AGN with $^{0.1}(u-g)>1.4$  is $50\pm4$ in the low  redshift subset of
the XMM/SDSS and $56\pm5$ in the AEGIS, where the errors are estimated
using binomial statistics.

\begin{figure}
\begin{center}
\includegraphics[height=0.9\columnwidth]{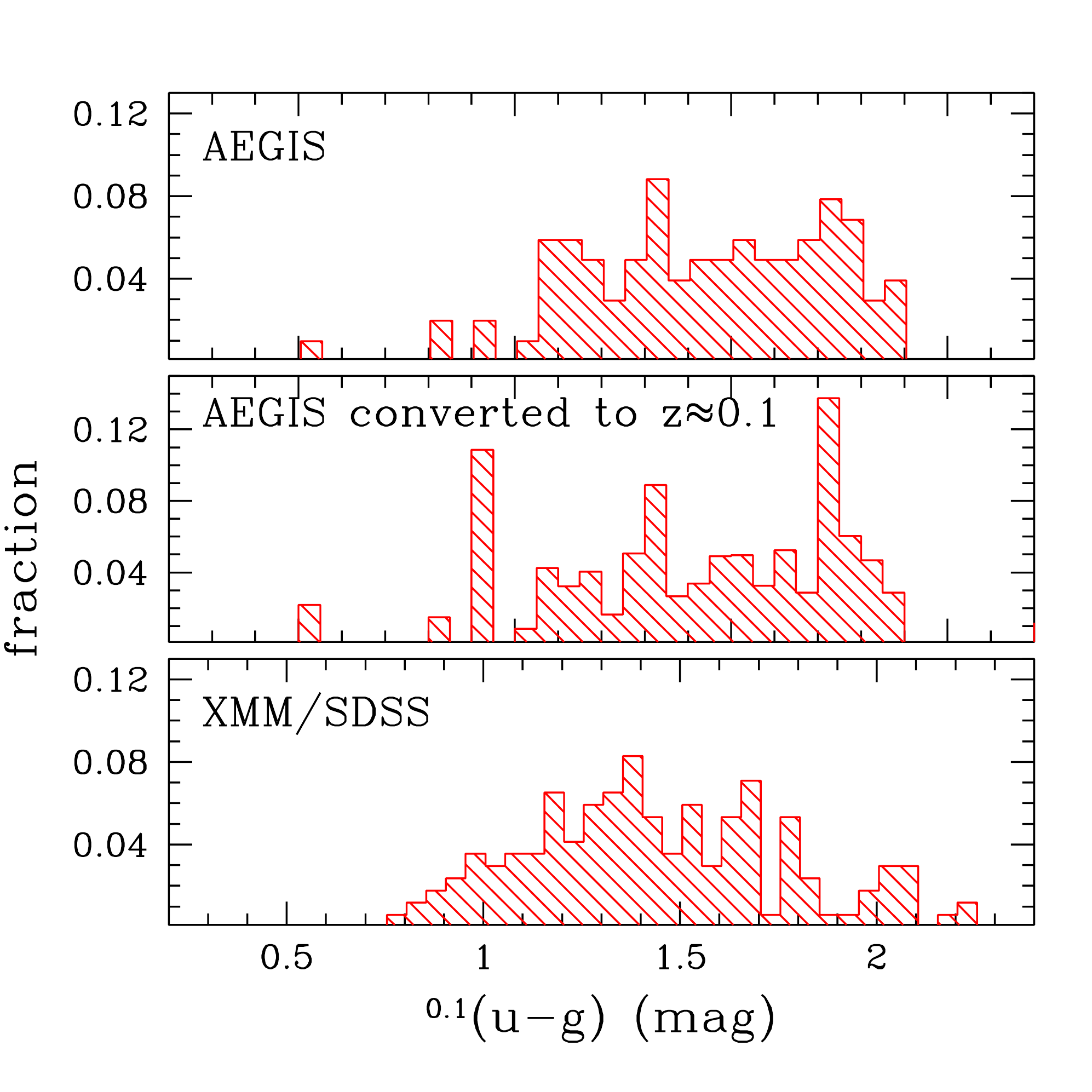}
\end{center}
\caption{The top panel shows  the $^{0.1}(u-g)$ colour distribution
  of X-ray AGN  at $0.6<z<1.2$  in the 
  AEGIS field. The middle panel shows the colour
  distribution of that sample as it would appear in the redshift interval
  $0.03<z<0.2$ in our XMM/SDSS survey. For comparison the bottom panel
  plots the  colour distribution  of X-ray AGN  in the range
  $0.03<z<0.2$ from the XMM/SDSS survey.}\label{fig_hist_cmd_sim}
\end{figure}

Before interpreting  those fractions  however, the differences  in the
selection function of the two samples need to be accounted for.  X-ray
AGN in  the AEGIS include, on  average, a higher fraction  of AGN with
high  hydrogen column  density,  $N_H> \rm  10^{22}  \, cm^{-2}$  (see
Figure \ref{fig_hist}  right panel).  The AGN light  of those obscured
AGN is less likely to affect  the host galaxy colours.  We account for
these  effects  using  the   simulations  described  in  the  appendix
\ref{sec_simulations}  to explore  how the  rest-frame colours  of the
AEGIS X-ray sources at $z\approx0.8$  would appear in the low redshift
($0.03<z<0.2$)  subsample of  the  XMM/SDSS survey.   In brief,  AEGIS
X-ray  sources are assigned  a random  redshift in  the range  $0.03 -
0.2$. Their  X-ray flux (2-10\,keV band) and  apparent magnitude (SDSS
$r$-band)  are estimated  at this  new redshift  to determine  if they
fulfill  the  selection   criteria  (e.g.   X-ray  sensitivity  curve,
$r<17.77$\,mag) of the low  redshift subsample of the XMM/SDSS survey.
We  can then  explore  how the  various  redshift dependent  selection
effects modify the derived  rest-frame colour distribution and if they
alter the fraction of AGN in red hosts.  The simulated distribution of
$^{0.1}(u-g)$ colours is plotted in Figure \ref{fig_hist_cmd_sim}.  It
is found that the fraction  of X-ray sources with $^{0.1}(u-g)>1.4$ is
55 per  cent, i.e.   the same as  for the $\approx0.8$  X-ray sources.
Differential selection effects are not important for the comparison of
the CMD  of X-ray  sources in  the AEGIS and  the XMM/SDSS  survey. We
therefore conclude that  the data are consistent with  no evolution of
the rest-frame colours of X-ray AGN from $z=0.8$ to $z=0.1$.

\section{Discussion and conclusions}

We present  a new serendipitous  X-ray survey, the XMM/SDSS,  which is
based on archival XMM observations  and covers $\rm 122\,deg^2$ in the
SDSS DR7 footprint.  The size  and sensitivity of this survey are well
suited  for  low redshift  X-ray  AGN  studies.   The XMM/SDSS  source
catalogue is  combined with the  SDSS optical spectroscopy  to compile
one of  the largest  hard X-ray (2-8\,keV)  AGN samples to date  in the
redshift interval $0.03<z<0.2$.  The selection function of this sample
is  well  defined and  similar  to  that  of deep,  pencil-beam  X-ray
surveys, which probe AGN at  moderate and high redshift.  This greatly
facilitates the study of the properties of AGN across redshift to shed
light on  the physical mechanism(s) responsible for  the rapid decline
of the accretion power of the Universe from $z\approx1$ to the present
day.   Variations with  redshift of  the dominant  SMBH  fueling mode,
which  in some  scenarios drive  the evolution  of the  AGN population
\citep[e.g.][]{Hopkins_Hernquist2006,   Hasinger2008,  Fanidakis2010},
are expected to  imprint on the properties of  AGN, including those of
their  host galaxies.   It has  been proposed  for example  that major
mergers  dominate  the  AGN  population  at high  redshift  and  bright
luminosities,  while  stochastic  accretion  onto  the  SMBH  of  disk
galaxies   become   important  at   low   redshift  and   luminosities
\citep{Hopkins_Hernquist2006}.  Alternatively, Fanidakis et al. (2010)
argue that disk  instabilities are the main AGN  fueling mode at early
epochs, while hot gas accretion becomes increasingly more important at
later times.

Systematic variations  with redshift of the rest-frame  colours of AGN
hosts  provide a  first order  test to  the scenarios  above.   In the
Fanidakis et  al. (2010) prescription for example,  the AGN population
should shift  to redder hosts with decreasing  redshift.  The opposite
trend might  be expected if disk instabilities  dominate the accretion
power at  later epochs. Although  a number of recent  studies explored
the   position  of   AGN   hosts  on   the  colour-magnitude   diagram
\citep[e.g.][]{Hickox2009, Cardamone2010,  Xue2010}, none of  them has
investigated  possible evolutionary  effects with  redshift, as  it is
done in this  paper.  We find no evidence for evolution  of the CMD of
X-ray  AGN   hosts  from  $z\approx0.8$   (AEGIS-X)  to  $z\approx0.1$
(XMM/SDSS).   Apart  from  changes   which  are  consistent  with  the
evolution of  the overall  galaxy population between  those redshifts,
X-ray AGN are  found in similar host galaxies,  in terms of integrated
colours, at both  $z=0.8$ and $z=0.1$.  This finding  does not support
the  scenario in  which the  dominant mode  of accretion  changes with
redshift since $z=0.8$.

It should  be noted however, that  a number of biases  may affect this
conclusion.  Although  we have removed sources  with colours dominated
by  the central  engine  (i.e. broad  optical  emission lines,  bright
nuclear point sources), we  cannot exclude the possibility of residual
contamination by scattered AGN light.  Nevertheless, \cite{Pierce2010}
studied  the HST colour  profiles of  AGN hosts  in the  AEGIS survey.
They concluded that the light from  the active nucleus does not have a
strong effect on the integrated colours  of the majority of the AGN in
their sample.  Recenty, \cite{Cardamone2010} suggest that about 25 per
cent of the red cloud  X-ray AGN at $z\approx1$ are dusty star-forming
systems, not  evolved galaxies with a dominant  old stellar population
component.   Systematic variations  with redshift  of the  fraction of
dust-reddened AGN hosts would impact our results and conclusions.

A striking  feature of Figure  \ref{fig_cmd} is the large  fraction of
blue  type-1 AGN  among low  redshift  X-ray sources  compared to  the
AEGIS-X  survey.  The  low  redshift XMM/SDSS  subsample inlcudes  XMM
targets, which in their majority are broad line AGN (13/16). Also, the
SDSS spectroscopic  targets inlcude apart from galaxies,  QSOs at both
low and high redshift, ROSAT X-ray sources, FIRST radio galaxies.  The
SDSS therefore has a higher spectroscopic sampling rate for type-1 AGN
compared to the  DEEP2 spectroscopic survey of the  AEGIS. We conclude
that selection effects are primarily responsible for the difference in
the relative fraction  of type-1 AGN begtween the  AEGIS-X and the low
redshift subset of the XMM/SDSS survey.

\section{Acknowledgments}
We thank  the anonymous referee for helpful  comments and suggestions.
The XMM/SDSS survey  data are available at www.astro.noa.gr/$\sim$age.
AG acknowledges  financial support from  the Marie-Curie Reintegration
Grant  PERG03-GA-2008-230644. Funding  for the  DEEP2  Galaxy Redshift
Survey  has   been  provided  in  part  by   NSF  grants  AST95-09298,
AST-0071048, AST-0071198, AST-0507428, and AST-0507483 as well as NASA
LTSA grant NNG04GC89G. Funding for the Sloan Digital Sky Survey (SDSS)
has   been  provided  by   the  Alfred   P.   Sloan   Foundation,  the
Participating  Institutions,   the  National  Aeronautics   and  Space
Administration, the National  Science Foundation, the U.S.  Department
of Energy,  the Japanese Monbukagakusho,  and the Max  Planck Society.
The SDSS Web site is http://www.sdss.org/.  The SDSS is managed by the
Astrophysical   Research  Consortium   (ARC)  for   the  Participating
Institutions.   The Participating Institutions  are The  University of
Chicago,  Fermilab,  the  Institute  for  Advanced  Study,  the  Japan
Participation Group, The Johns Hopkins University, Los Alamos National
Laboratory,  the   Max-Planck-Institute  for  Astronomy   (MPIA),  the
Max-Planck-Institute   for  Astrophysics   (MPA),  New   Mexico  State
University, University of Pittsburgh, Princeton University, the United
States Naval Observatory, and the University of Washington.

\bibliography{mybib}{}

\begin{thebibliography}{67}
\expandafter\ifx\csname natexlab\endcsname\relax\def\natexlab#1{#1}\fi

\bibitem[{{Abazajian} {et~al.}(2009)}]{Abazajian2009}
{Abazajian} K.~N., {et~al.}, 2009, ApJS, 182, 543

\bibitem[{{Aird} {et~al.}(2010)}]{Aird2010}
{Aird} J., {et~al.}, 2010, MNRAS, 401, 2531

\bibitem[{{Barnes} \& {Hernquist}(1996)}]{Barnes1996}
{Barnes} J.~E., {Hernquist} L., 1996, ApJ, 471, 115

\bibitem[{{Barnes} \& {Hernquist}(1991)}]{Barnes1991}
{Barnes} J.~E., {Hernquist} L.~E., 1991, ApJ, 370, L65

\bibitem[{{Becker} {et~al.}(1995){Becker}, {White}, \& {Helfand}}]{Becker1995}
{Becker} R.~H., {White} R.~L., {Helfand} D.~J., 1995, ApJ, 450, 559

\bibitem[{{Beckmann} {et~al.}(2009){Beckmann}, {Soldi}, {Ricci},
  {Alfonso-Garz{\'o}n}, {Courvoisier}, {Domingo}, {Gehrels}, {Lubi{\'n}ski},
  {Mas-Hesse}, \& {Zdziarski}}]{Beckmann2009}
{Beckmann} V., {Soldi} S., {Ricci} C., {Alfonso-Garz{\'o}n} J., {Courvoisier}
  T., {Domingo} A., {Gehrels} N., {Lubi{\'n}ski} P., {Mas-Hesse} J.~M.,
  {Zdziarski} A.~A., 2009, A\&A, 505, 417

\bibitem[{{Bell} {et~al.}(2004){Bell}, {Wolf}, {Meisenheimer}, {Rix}, {Borch},
  {Dye}, {Kleinheinrich}, {Wisotzki}, \& {McIntosh}}]{Bell2004}
{Bell} E.~F., {Wolf} C., {Meisenheimer} K., {Rix} H., {Borch} A., {Dye} S.,
  {Kleinheinrich} M., {Wisotzki} L., {McIntosh} D.~H., 2004, ApJ, 608, 752

\bibitem[{{Blanton}(2006)}]{Blanton2006}
{Blanton} M.~R., 2006, ApJ, 648, 268

\bibitem[{{Blanton} \& {Roweis}(2007)}]{Blanton_Roweis2007}
{Blanton} M.~R., {Roweis} S., 2007, AJ, 133, 734

\bibitem[{{Blanton} {et~al.}(2005){Blanton}, {Schlegel}, {Strauss},
  {Brinkmann}, {Finkbeiner}, {Fukugita}, {Gunn}, {Hogg}, {Ivezi{\'c}}, {Knapp},
  {Lupton}, {Munn}, {Schneider}, {Tegmark}, \& {Zehavi}}]{Blanton2005}
{Blanton} M.~R., {Schlegel} D.~J., {Strauss} M.~A., {Brinkmann} J.,
  {Finkbeiner} D., {Fukugita} M., {Gunn} J.~E., {Hogg} D.~W., {Ivezi{\'c}} {\v
  Z}., {Knapp} G.~R., {Lupton} R.~H., {Munn} J.~A., {Schneider} D.~P.,
  {Tegmark} M., {Zehavi} I., 2005, AJ, 129, 2562

\bibitem[{{Cardamone} {et~al.}(2010){Cardamone}, {Urry}, {Schawinski},
  {Treister}, {Brammer}, \& {Gawiser}}]{Cardamone2010}
{Cardamone} C.~N., {Urry} C.~M., {Schawinski} K., {Treister} E., {Brammer} G.,
  {Gawiser} E., 2010, ApJ, 721, L38

\bibitem[{{Cattaneo} {et~al.}(2005){Cattaneo}, {Blaizot}, {Devriendt}, \&
  {Guiderdoni}}]{Cattaneo2005}
{Cattaneo} A., {Blaizot} J., {Devriendt} J., {Guiderdoni} B., 2005, MNRAS, 364,
  407

\bibitem[{{Coil} {et~al.}(2004){Coil}, {Newman}, {Kaiser}, {Davis}, {Ma},
  {Kocevski}, \& {Koo}}]{Coil2004}
{Coil} A.~L., {Newman} J.~A., {Kaiser} N., {Davis} M., {Ma} C., {Kocevski}
  D.~D., {Koo} D.~C., 2004, ApJ, 617, 765

\bibitem[{{Croton} {et~al.}(2006)}]{Croton2006}
{Croton} D.~J., {et~al.}, 2006, MNRAS, 365, 11

\bibitem[{{Davis} {et~al.}(2007)}]{Davis2007}
{Davis} M., {et~al.}, 2007, ApJ, 660, L1

\bibitem[{{Di Matteo} {et~al.}(2005){Di Matteo}, {Springel}, \&
  {Hernquist}}]{DiMatteo2005}
{Di Matteo} T., {Springel} V., {Hernquist} L., 2005, Nature, 433, 604

\bibitem[{{Donley} {et~al.}(2008){Donley}, {Rieke}, {P{\'e}rez-Gonz{\'a}lez},
  \& {Barro}}]{Donley2008}
{Donley} J.~L., {Rieke} G.~H., {P{\'e}rez-Gonz{\'a}lez} P.~G., {Barro} G.,
  2008, ApJ, 687, 111

\bibitem[{{Eales} {et~al.}(2010)}]{Eales2010}
{Eales} S., {et~al.}, 2010, PASP, 122, 499

\bibitem[{{Ebrero} {et~al.}(2009){Ebrero}, {Carrera}, {Page}, {Silverman},
  {Barcons}, {Ceballos}, {Corral}, {Della Ceca}, \& {Watson}}]{Ebrero2009}
{Ebrero} J., {Carrera} F.~J., {Page} M.~J., {Silverman} J.~D., {Barcons} X.,
  {Ceballos} M.~T., {Corral} A., {Della Ceca} R., {Watson} M.~G., 2009, A\&A,
  493, 55

\bibitem[{{Fanidakis} {et~al.}(2010){Fanidakis}, {Baugh}, {Benson}, {Bower},
  {Cole}, {Done}, {Frenk}, {Hickox}, {Lacey}, \& {Lagos}}]{Fanidakis2010}
{Fanidakis} N., {Baugh} C.~M., {Benson} A.~J., {Bower} R.~G., {Cole} S., {Done}
  C., {Frenk} C.~S., {Hickox} R.~C., {Lacey} C., {Lagos} C.~d.~P., 2010,
  arXiv-1011.5222

\bibitem[{{Ferrarese} \& {Merritt}(2000)}]{Ferrarese2000}
{Ferrarese} L., {Merritt} D., 2000, ApJ, 539, L9

\bibitem[{{Gebhardt} {et~al.}(2000){Gebhardt}, {Bender}, {Bower}, {Dressler},
  {Faber}, {Filippenko}, {Green}, {Grillmair}, {Ho}, {Kormendy}, {Lauer},
  {Magorrian}, {Pinkney}, {Richstone}, \& {Tremaine}}]{Gebhardt2000}
{Gebhardt} K., {Bender} R., {Bower} G., {Dressler} A., {Faber} S.~M.,
  {Filippenko} A.~V., {Green} R., {Grillmair} C., {Ho} L.~C., {Kormendy} J.,
  {Lauer} T.~R., {Magorrian} J., {Pinkney} J., {Richstone} D., {Tremaine} S.,
  2000, ApJ, 539, L13

\bibitem[{{Georgakakis}(2008)}]{Georgakakis2008AN}
{Georgakakis} A., 2008, Astronomische Nachrichten, 329, 174

\bibitem[{{Georgakakis} {et~al.}(2008){Georgakakis}, {Nandra}, {Laird}, {Aird},
  \& {Trichas}}]{Georgakakis2008_sense}
{Georgakakis} A., {Nandra} K., {Laird} E.~S., {Aird} J., {Trichas} M., 2008,
  MNRAS, 388, 1205

\bibitem[{{Georgakakis} {et~al.}(2007){Georgakakis}, {Rowan-Robinson},
  {Babbedge}, \& {Georgantopoulos}}]{Georgakakis2007}
{Georgakakis} A., {Rowan-Robinson} M., {Babbedge} T.~S.~R., {Georgantopoulos}
  I., 2007, MNRAS, 377, 203

\bibitem[{{Georgakakis} {et~al.}(2010){Georgakakis}, {Rowan-Robinson},
  {Nandra}, {Digby-North}, {P{\'e}rez-Gonz{\'a}lez}, \&
  {Barro}}]{Georgakakis2010}
{Georgakakis} A., {Rowan-Robinson} M., {Nandra} K., {Digby-North} J.,
  {P{\'e}rez-Gonz{\'a}lez} P.~G., {Barro} G., 2010, MNRAS, 406, 420

\bibitem[{{Georgantopoulos} {et~al.}(2008){Georgantopoulos}, {Georgakakis},
  {Rowan-Robinson}, \& {Rovilos}}]{Georgantopoulos2008}
{Georgantopoulos} I., {Georgakakis} A., {Rowan-Robinson} M., {Rovilos} E.,
  2008, A\&A, 484, 671

\bibitem[{{Hamilton} \& {Tegmark}(2004)}]{Hamilton_Tegmark2004}
{Hamilton} A.~J.~S., {Tegmark} M., 2004, MNRAS, 349, 115

\bibitem[{{Hasinger}(2008)}]{Hasinger2008}
{Hasinger} G., 2008, A\&A, 490, 905

\bibitem[{{Hasinger} {et~al.}(2005){Hasinger}, {Miyaji}, \&
  {Schmidt}}]{Hasinger2005}
{Hasinger} G., {Miyaji} T., {Schmidt} M., 2005, A\&A, 441, 417

\bibitem[{{Hernquist}(1989)}]{Hernquist1989}
{Hernquist} L., 1989, Nature, 340, 687

\bibitem[{{Hickox} \& {Markevitch}(2006)}]{Hickox2006}
{Hickox} R.~C., {Markevitch} M., 2006, ApJ, 645, 95

\bibitem[{{Hickox} {et~al.}(2009)}]{Hickox2009}
{Hickox} R.~C., {et~al.}, 2009, ArXiv0901.4121

\bibitem[{{Hopkins} \& {Hernquist}(2006)}]{Hopkins_Hernquist2006}
{Hopkins} P.~F., {Hernquist} L., 2006, ApJS, 166, 1

\bibitem[{{Hopkins} {et~al.}(2006){Hopkins}, {Hernquist}, {Cox}, {Di Matteo},
  {Robertson}, \& {Springel}}]{Hopkins2006}
{Hopkins} P.~F., {Hernquist} L., {Cox} T.~J., {Di Matteo} T., {Robertson} B.,
  {Springel} V., 2006, ApJS, 163, 1

\bibitem[{{Ishihara} {et~al.}(2010)}]{Ishihara2010}
{Ishihara} D., {et~al.}, 2010, A\&A, 514, A1+

\bibitem[{{Kalberla} {et~al.}(2005){Kalberla}, {Burton}, {Hartmann}, {Arnal},
  {Bajaja}, {Morras}, \& {P{\"o}ppel}}]{Kalberla2005}
{Kalberla} P.~M.~W., {Burton} W.~B., {Hartmann} D., {Arnal} E.~M., {Bajaja} E.,
  {Morras} R., {P{\"o}ppel} W.~G.~L., 2005, A\&A, 440, 775

\bibitem[{{Kauffmann} {et~al.}(2004){Kauffmann}, {White}, {Heckman},
  {M{\'e}nard}, {Brinchmann}, {Charlot}, {Tremonti}, \&
  {Brinkmann}}]{Kauffmann2004}
{Kauffmann} G., {White} S.~D.~M., {Heckman} T.~M., {M{\'e}nard} B.,
  {Brinchmann} J., {Charlot} S., {Tremonti} C., {Brinkmann} J., 2004, MNRAS,
  353, 713

\bibitem[{{Laird} {et~al.}(2009)}]{Laird2009}
{Laird} E.~S., {et~al.}, 2009, ApJS, 180, 102

\bibitem[{{Lawrence} {et~al.}(2007)}]{Lawrence2007}
{Lawrence} A., {et~al.}, 2007, MNRAS, 379, 1599

\bibitem[{{L{\'o}pez-Sanjuan} {et~al.}(2009){L{\'o}pez-Sanjuan}, {Balcells},
  {P{\'e}rez-Gonz{\'a}lez}, {Barro}, {Garc{\'{\i}}a-Dab{\'o}}, {Gallego}, \&
  {Zamorano}}]{LopezSanjuan2009}
{L{\'o}pez-Sanjuan} C., {Balcells} M., {P{\'e}rez-Gonz{\'a}lez} P.~G., {Barro}
  G., {Garc{\'{\i}}a-Dab{\'o}} C.~E., {Gallego} J., {Zamorano} J., 2009, A\&A,
  501, 505

\bibitem[{{Lumb} {et~al.}(2002){Lumb}, {Warwick}, {Page}, \& {De
  Luca}}]{Lumb2002}
{Lumb} D.~H., {Warwick} R.~S., {Page} M., {De Luca} A., 2002, A\&A, 389, 93

\bibitem[{{Marty} {et~al.}(2003){Marty}, {Kneib}, {Sadat}, {Ebeling}, \&
  {Smail}}]{Marty2003}
{Marty} P.~B., {Kneib} J., {Sadat} R., {Ebeling} H., {Smail} I., 2003, in
  Society of Photo-Optical Instrumentation Engineers (SPIE) Conference Series,
  Vol. 4851, Society of Photo-Optical Instrumentation Engineers (SPIE)
  Conference Series, {J.~E.~Truemper \& H.~D.~Tananbaum}, ed., pp. 208--222

\bibitem[{{Morrison} \& {McCammon}(1983)}]{Morrison1983}
{Morrison} R., {McCammon} D., 1983, ApJ, 270, 119

\bibitem[{{Nandra} {et~al.}(2005){Nandra}, {Laird}, {Adelberger}, {Gardner},
  {Mushotzky}, {Rhodes}, {Steidel}, {Teplitz}, \& {Arnaud}}]{Nandra2005}
{Nandra} K., {Laird} E.~S., {Adelberger} K., {Gardner} J.~P., {Mushotzky}
  R.~F., {Rhodes} J., {Steidel} C.~C., {Teplitz} H.~I., {Arnaud} K.~A., 2005,
  MNRAS, 356, 568

\bibitem[{{Nandra} {et~al.}(2007){Nandra}, {O'Neill}, {George}, \&
  {Reeves}}]{Nandra2007_Fe}
{Nandra} K., {O'Neill} P.~M., {George} I.~M., {Reeves} J.~N., 2007, MNRAS, 382,
  194

\bibitem[{{Nandra} \& {Pounds}(1994)}]{Nandra1994}
{Nandra} K., {Pounds} K.~A., 1994, MNRAS, 268, 405

\bibitem[{{Oyaizu} {et~al.}(2008){Oyaizu}, {Lima}, {Cunha}, {Lin}, {Frieman},
  \& {Sheldon}}]{Oyaizu2008}
{Oyaizu} H., {Lima} M., {Cunha} C.~E., {Lin} H., {Frieman} J., {Sheldon} E.~S.,
  2008, ApJ, 674, 768

\bibitem[{{Padmanabhan} {et~al.}(2008)}]{Padmanabhan2008}
{Padmanabhan} N., {et~al.}, 2008, ApJ, 674, 1217

\bibitem[{{Park} {et~al.}(2006){Park}, {Kashyap}, {Siemiginowska}, {van Dyk},
  {Zezas}, {Heinke}, \& {Wargelin}}]{Park2006}
{Park} T., {Kashyap} V.~L., {Siemiginowska} A., {van Dyk} D.~A., {Zezas} A.,
  {Heinke} C., {Wargelin} B.~J., 2006, ApJ, 652, 610

\bibitem[{{Pier} {et~al.}(2003){Pier}, {Munn}, {Hindsley}, {Hennessy}, {Kent},
  {Lupton}, \& {Ivezi{\'c}}}]{Pier2003}
{Pier} J.~R., {Munn} J.~A., {Hindsley} R.~B., {Hennessy} G.~S., {Kent} S.~M.,
  {Lupton} R.~H., {Ivezi{\'c}} {\v Z}., 2003, AJ, 125, 1559

\bibitem[{{Pierce} {et~al.}(2010)}]{Pierce2010}
{Pierce} C.~M., {et~al.}, 2010, MNRAS, 408, 139

\bibitem[{{Pope} {et~al.}(2008)}]{Pope2008}
{Pope} A., {et~al.}, 2008, ApJ, 689, 127

\bibitem[{{Sarzi} {et~al.}(2010)}]{Sarzi2010}
{Sarzi} M., {et~al.}, 2010, MNRAS, 402, 2187

\bibitem[{{Schlegel} {et~al.}(1998){Schlegel}, {Finkbeiner}, \&
  {Davis}}]{Schlegel1998}
{Schlegel} D.~J., {Finkbeiner} D.~P., {Davis} M., 1998, ApJ, 500, 525

\bibitem[{{Severgnini} {et~al.}(2003)}]{Severgnini2003}
{Severgnini} P., {et~al.}, 2003, A\&A, 406, 483

\bibitem[{{Somerville} {et~al.}(2008){Somerville}, {Hopkins}, {Cox},
  {Robertson}, \& {Hernquist}}]{Somerville2008}
{Somerville} R.~S., {Hopkins} P.~F., {Cox} T.~J., {Robertson} B.~E.,
  {Hernquist} L., 2008, ArXiv0808.1227

\bibitem[{{Springel} {et~al.}(2005)}]{Springel2005}
{Springel} V., {et~al.}, 2005, Nature, 435, 629

\bibitem[{{Strauss} {et~al.}(2002)}]{Strauss2002}
{Strauss} M.~A., {et~al.}, 2002, AJ, 124, 1810

\bibitem[{{Sutherland} \& {Saunders}(1992)}]{Sutherland_and_Saunders1992}
{Sutherland} W., {Saunders} W., 1992, MNRAS, 259, 413

\bibitem[{{Tueller} {et~al.}(2010)}]{Tueller2010}
{Tueller} J., {et~al.}, 2010, ApJS, 186, 378

\bibitem[{{Ueda} {et~al.}(2003){Ueda}, {Akiyama}, {Ohta}, \&
  {Miyaji}}]{Ueda2003}
{Ueda} Y., {Akiyama} M., {Ohta} K., {Miyaji} T., 2003, ApJ, 598, 886

\bibitem[{{Wang} {et~al.}(2008){Wang}, {De Lucia}, {Kitzbichler}, \&
  {White}}]{Wang2008}
{Wang} J., {De Lucia} G., {Kitzbichler} M.~G., {White} S.~D.~M., 2008, MNRAS,
  384, 1301

\bibitem[{{Watson} {et~al.}(2009)}]{Watson2009}
{Watson} M.~G., {et~al.}, 2009, A\&A, 493, 339

\bibitem[{{Weiner} {et~al.}(2005)}]{Weiner2005}
{Weiner} B.~J., {et~al.}, 2005, ApJ, 620, 595

\bibitem[{{Willmer} {et~al.}(2006)}]{Willmer2006}
{Willmer} C.~N.~A., {et~al.}, 2006, ApJ, 647, 853

\bibitem[{{Xue} {et~al.}(2010){Xue}, {Brandt}, {Luo}, {Rafferty}, {Alexander},
  {Bauer}, {Lehmer}, {Schneider}, \& {Silverman}}]{Xue2010}
{Xue} Y.~Q., {Brandt} W.~N., {Luo} B., {Rafferty} D.~A., {Alexander} D.~M.,
  {Bauer} F.~E., {Lehmer} B.~D., {Schneider} D.~P., {Silverman} J.~D., 2010,
  ApJ, 720, 368

\end{thebibliography}
\bibliographystyle{mn2e}

\appendix
\section{Point Spread Function}\label{sec_psf}
The XMM  Science Simulator  ({\sc SciSim}) version  4.0.4 was  used to
produce     point    sources     with     monochromatic    flux     of
$2\times10^{-5}$\,photons/mm/s at 1, 2,  3.5, 6 and 8.5\,keV, i.e. the
mean  energies  of   the  soft,  full,  hard,  uhrd   and  vhrd  bands
respectively. The simulated sources were placed in a grid of positions
separated by  90\,arcsec on the PN,  MOS1 and MOS2  cameras.  A denser
grid, although  desirable, would be  computationally expensive.  Also,
no  simulations  were produced  for  positions  that  are closer  than
21\,arcsec radius  from CCD gaps, as  in those cases the  PSF size and
shape cannot be reliably determined.   The total exposure time was set
to 50\,ks.   Because of technical  issues related to the  {\sc SciSim}
code, this  was split into  5 separate 10\,ks simulations,  which were
then merged into  a single event file. The source  photon flux and the
integration time were chosen to produce sufficient number of counts on
the  EPIC cameras to  determine the  PSF properties.   Each simulation
consists of a single energy, position and detector. There are 324, 361
and  380  PSFs at  each  energy  for the  PN,  MOS1  and MOS2  cameras
respectively.

The {\sc SciSim}  generated ODF level files which  were processed with
SAS  to  produce  event  files,  images and  exposure  maps  for  each
simulation.  The source  counts were fit with ellipses  using the {\sc
  IRAF} (Image Reduction and  Analysis Facility) task {\sc ellipse} of
the {\sc stsdas}  package, which was designed to  fit the isophotes of
optical  galaxies.   The  task  reads 2-dimensional  images  and  fits
ellipses to the isophotes at predefined semi-major axis values. During
the  ellipse  fitting  process  the  centre of  the  source  was  kept
fixed. The position angle,  PA, and ellipticity, $\epsilon$, were left
free parameters up to the  semi-major axis value where S/N constraints
terminated  the fitting.  Beyond  that radius  the position  angle and
ellipticity were kept constant.  Parts of the image with exposure time
equal  to zero  (e.g. CCD  gaps) were  masked out  during  the fitting
process. The output of the task  is a table that includes, among other
parameters, the  best fit ellipticity and position  angle at different
semi-major axis values.  We used  those parameters to estimate the PSF
profile by  extracting counts within elliptical  annuli.  The fraction
of the pixels of the annuli that fall either outside the EPIC field of
view  or within CCD  gaps is  accounted for  in the  calculations. The
cumulative radial distribution of the corrected counts is constructed,
which is then  used to estimate by linear  interpolation the semi major
axis radius, $\epsilon$  and PA of the  60, 70 and 80 per  cent EEF of
the PSF at different monochromatic  energies and positions on the EPIC
CCDs.

\section{Table columns}\label{sec_key}

\subsection{source detection parameters}

UXID (30A): Unique X-ray source identification string. The unique name
of the  each source  is a  string which is  composed from  the project
name, followed  by the number of  the pointing on which  the source is
detected or zero in the case of projects with a single XMM observation
and ending with a sequential number.

\noindent 
BOX\_ID\_SRC  (J): identification  number  of a  source on  individual
pointings or  observations in the case  of projects with  a single XMM
observation.

\noindent 
RA  (D):  Right  Ascension  in  degrees of  the  X-ray  source  before
correcting for systematic offsets.

\noindent 
DEC (D): Declination in degrees  of the X-ray source before correcting
for systematic offsets.

\noindent 
RADEC\_ERR  (E):  Positional  uncertainty  in  arcsec  of  the  source
position. This parameter  is estimated by the {\sc  ewavelet} task of SAS. 

\noindent 
CNT\_TOTAL\_FULL\_70 (J): full band counts at the source position from
all EPIC cameras within the 70 per cent EEF ellipse.

\noindent 
BKG\_TOTAL\_FULL\_70  (E):  expected number  of  full band  background
counts at the source position from  all EPIC cameras within the 70 per
cent EEF ellipse.

\noindent 
EXP\_TOTAL\_FULL\_70 (E):  full band exposure  time in seconds  at the
source position from  all EPIC cameras. This parameter  is the average
exposure time within the 70 per cent EEF ellipse.

\noindent 
PROB\_FULL (E): Poisson probability that the source is spurious in the
full band.

\noindent 
CNT\_PN\_FULL\_70 (J): full band counts  at the source position on the
PN camera within the 70 per cent EEF ellipse.

\noindent 
BKG\_PN\_FULL\_70 (E): expected number  of full band background counts
at the  source position on  the PN camera  within the 70 per  cent EEF
ellipse.

\noindent 
EXP\_PN\_FULL\_70  (E): PN  camera  exposure time  in  seconds at  the
source position.   This parameter is the average  exposure time within
the 70 per cent EEF ellipse.

\noindent 
CNT\_M1\_FULL\_70 (J): full band counts  at the source position on the
MOS1 camera within the 70 per cent EEF ellipse.

\noindent 
BKG\_M1\_FULL\_70 (E): expected number  of full band background counts
at the source  position on the MOS1 camera within the  70 per cent EEF
ellipse.

\noindent 
EXP\_M1\_FULL\_70  (E): MOS1 camera  exposure time  in seconds  at the
source position.   This parameter is the average  exposure time within
the 70 per cent EEF ellipse.

\noindent 
CNT\_M2\_FULL\_70 (J): full band counts  at the source position on the
MOS2 camera within the 70 per cent EEF ellipse.

\noindent 
BKG\_M2\_FULL\_70 (E): expected number  of full band background counts
at the source  position on the MOS2 camera within the  70 per cent EEF
ellipse.

\noindent 
EXP\_M2\_FULL\_70  (E): MOS2 camera  exposure time  in seconds  at the
source position.   This parameter is the average  exposure time within
the 70 per cent EEF ellipse.

\noindent 
EML\_DET\_ML\_FULL  (E): {\sc emldetect}  detection likelihood  in the
full band for all EPIC cameras.

\noindent 
EML\_EXT\_ML\_FULL (E): Likelihood that  the source is extended in the
full band. This parameter is  estimated by the {\sc emldetect} task of
SAS for all EPIC cameras.

\noindent 
CNT\_TOTAL\_SOFT\_70 (J): soft band counts at the source position from
all EPIC cameras within the 70 per cent EEF ellipse.

\noindent 
BKG\_TOTAL\_SOFT\_70  (E):  expected number  of  soft band  background
counts at the source position from  all EPIC cameras within the 70 per
cent EEF ellipse.

\noindent 
EXP\_TOTAL\_SOFT\_70 (E):  soft band exposure  time in seconds  at the
source position from  all EPIC cameras. This parameter  is the average
exposure time within the 70 per cent EEF ellipse.

\noindent 
PROB\_SOFT (E): Poisson probability that the source is spurious in the
soft band.

\noindent 
CNT\_PN\_SOFT\_70 (J): soft  band counts at the source  position on the
PN camera within the 70 per cent EEF ellipse.

\noindent 
BKG\_PN\_SOFT\_70 (E): expected number  of soft band background counts
at the  source position on  the PN camera  within the 70 per  cent EEF
ellipse.

\noindent 
EXP\_PN\_SOFT\_70 (E): PN camera soft-band exposure time in seconds at
the  source position.   This parameter  is the  average  exposure time
within the 70 per cent EEF ellipse.

\noindent 
CNT\_M1\_SOFT\_70 (J): soft band counts  at the source position on the
MOS1 camera within the 70 per cent EEF ellipse.

\noindent 
BKG\_M1\_SOFT\_70 (E): expected number  of soft band background counts
at the source  position on the MOS1 camera within the  70 per cent EEF
ellipse.

\noindent 
EXP\_M1\_SOFT\_70 (E): MOS1 camera  soft-band exposure time in seconds
at the source  position.  This parameter is the  average exposure time
within the 70 per cent EEF ellipse.

\noindent 
CNT\_M2\_SOFT\_70 (J): soft band counts  at the source position on the
MOS2 camera within the 70 per cent EEF ellipse.

\noindent 
BKG\_M2\_SOFT\_70 (E): expected number  of soft band background counts
at the source  position on the MOS2 camera within the  70 per cent EEF
ellipse.

\noindent 
EXP\_M2\_SOFT\_70 (E): MOS2 camera  soft-band exposure time in seconds
at the source  position.  This parameter is the  average exposure time
within the 70 per cent EEF ellipse.
 
\noindent 
EML\_DET\_ML\_SOFT  (E): {\sc emldetect}  detection likelihood  in the
soft band for all EPIC cameras.

\noindent 
EML\_EXT\_ML\_SOFT (E): Likelihood that  the source is extended in the
soft band. This parameter is  estimated by the {\sc emldetect} task of
SAS for all EPIC cameras.

\noindent 
CNT\_TOTAL\_HARD\_70 (J): hard band counts at the source position from
all EPIC cameras within the 70 per cent EEF ellipse.

\noindent 
BKG\_TOTAL\_HARD\_70  (E):  expected number  of  hard band  background
counts at the source position from  all EPIC cameras within the 70 per
cent EEF ellipse.

\noindent 
EXP\_TOTAL\_HARD\_70 (E):  hard band exposure  time in seconds  at the
source position from  all EPIC cameras. This parameter  is the average
exposure time within the 70 per cent EEF ellipse.

\noindent 
PROB\_HARD (E): Poisson probability that the source is spurious in the
hard band.

\noindent 
CNT\_PN\_HARD\_70 (J): hard band counts  at the source position on the
PN camera within the 70 per cent EEF ellipse.

\noindent 
BKG\_PN\_HARD\_70 (E): expected number  of hard band background counts
at the  source position on  the PN camera  within the 70 per  cent EEF
ellipse.

\noindent 
EXP\_PN\_HARD\_70 (E): PN camera hard-band exposure time in seconds at
the  source position.   This parameter  is the  average  exposure time
within the 70 per cent EEF ellipse.

\noindent 
CNT\_M1\_HARD\_70 (J): hard band counts  at the source position on the
MOS1 camera within the 70 per cent EEF ellipse.

\noindent 
BKG\_M1\_HARD\_70 (E): expected number  of hard band background counts
at the source  position on the MOS1 camera within the  70 per cent EEF
ellipse.

\noindent 
EXP\_M1\_HARD\_70 (E): MOS1 camera  hard-band exposure time in seconds
at the source  position.  This parameter is the  average exposure time
within the 70 per cent EEF ellipse.

\noindent 
CNT\_M2\_HARD\_70 (J): hard band counts  at the source position on the
MOS2 camera within the 70 per cent EEF ellipse.

\noindent 
BKG\_M2\_HARD\_70 (E): expected number  of hard band background counts
at the source  position on the MOS2 camera within the  70 per cent EEF
ellipse.

\noindent 
EXP\_M2\_HARD\_70 (E): MOS2 camera  hard-band exposure time in seconds
at the source  position.  This parameter is the  average exposure time
within the 70 per cent EEF ellipse.

\noindent 
EML\_DET\_ML\_HARD  (E): {\sc emldetect}  detection likelihood  in the
hard band for all EPIC cameras.

\noindent 
EML\_EXT\_ML\_HARD (E): Likelihood that  the source is extended in the
hard band. This parameter is  estimated by the {\sc emldetect} task of
SAS for all EPIC cameras.

\noindent 
CNT\_TOTAL\_VHRD\_70 (J): very hard band counts at the source position
from all EPIC cameras within the 70 per cent EEF ellipse.

\noindent 
BKG\_TOTAL\_VHRD\_70 (E): expected number of very hard band background
counts at the source position from  all EPIC cameras within the 70 per
cent EEF ellipse.

\noindent 
EXP\_TOTAL\_VHRD\_70 (E):  very hard band exposure time  in seconds at
the  source position  from all  EPIC  cameras. This  parameter is  the
average exposure time within the 70 per cent EEF ellipse.

\noindent 
PROB\_VHRD (E): Poisson probability that the source is spurious in the
very hard band.

\noindent 
CNT\_PN\_VHRD\_70 (J): very hard band counts at the source position on
the PN camera within the 70 per cent EEF ellipse.

\noindent 
BKG\_PN\_VHRD\_70 (E):  expected number  of very hard  band background
counts at the source position on  the PN camera within the 70 per cent
EEF ellipse.

\noindent 
EXP\_PN\_VHRD\_70  (E): PN  camera  very hard  band  exposure time  in
seconds  at  the  source  position.   This parameter  is  the  average
exposure time within the 70 per cent EEF ellipse.

\noindent 
CNT\_M1\_VHRD\_70 (J): very hard band counts at the source position on
the MOS1 camera within the 70 per cent EEF ellipse.

\noindent 
BKG\_M1\_VHRD\_70 (E):  expected number  of very hard  band background
counts at  the source position  on the MOS1  camera within the  70 per
cent EEF ellipse.

\noindent 
EXP\_M1\_VHRD\_70  (E): MOS1 camera  very hard  band exposure  time in
seconds  at  the  source  position.   This parameter  is  the  average
exposure time within the 70 per cent EEF ellipse.

\noindent 
CNT\_M2\_VHRD\_70 (J): very hard band counts at the source position on
the MOS2 camera within the 70 per cent EEF ellipse.

\noindent 
BKG\_M2\_VHRD\_70 (E):  expected number  of very hard  band background
counts at  the source position  on the MOS2  camera within the  70 per
cent EEF ellipse.

\noindent 
EXP\_M2\_VHRD\_70  (E): MOS2 camera  very hard  band exposure  time in
seconds  at  the  source  position.   This parameter  is  the  average
exposure time within the 70 per cent EEF ellipse.

\noindent 
EML\_DET\_ML\_VHRD  (E): {\sc emldetect}  detection likelihood  in the
very hard band for all EPIC cameras.

\noindent 
EML\_EXT\_ML\_VHRD (E): Likelihood that  the source is extended in the
very hard  band. This  parameter is estimated  by the  {\sc emldetect}
task of SAS for all EPIC cameras.

\noindent 
CNT\_TOTAL\_UHRD\_70 (J): ultra hard band counts at the source position
from all EPIC cameras within the 70 per cent EEF ellipse.

\noindent 
BKG\_TOTAL\_UHRD\_70   (E):  expected  number   of  ultra   hard  band
background counts at the source  position from all EPIC cameras within
the 70 per cent EEF ellipse.

\noindent 
EXP\_TOTAL\_UHRD\_70 (E): ultra hard  band exposure time in seconds at
the  source position  from all  EPIC  cameras. This  parameter is  the
average exposure time within the 70 per cent EEF ellipse.

\noindent 
PROB\_UHRD (E): Poisson probability that the source is spurious in the
ultra hard band.

\noindent 
CNT\_PN\_UHRD\_70 (J):  ultra hard band counts at  the source position
on the PN camera within the 70 per cent EEF ellipse.

\noindent 
BKG\_PN\_UHRD\_70 (E):  expected number of ultra  hard band background
counts at the source position on  the PN camera within the 70 per cent
EEF ellipse.

\noindent 
EXP\_PN\_UHRD\_70  (E): PN  camera ultra  hard band  exposure  time in
seconds  at  the  source  position.   This parameter  is  the  average
exposure time within the 70 per cent EEF ellipse.

\noindent 
CNT\_M1\_UHRD\_70 (J):  ultra hard band counts at  the source position
on the MOS1 camera within the 70 per cent EEF ellipse.

\noindent 
BKG\_M1\_UHRD\_70 (E):  expected number of ultra  hard band background
counts at  the source position  on the MOS1  camera within the  70 per
cent EEF ellipse.

\noindent 
EXP\_M1\_UHRD\_70 (E):  MOS1 camera ultra  hard band exposure  time in
seconds  at  the  source  position.   This parameter  is  the  average
exposure time within the 70 per cent EEF ellipse.

\noindent 
CNT\_M2\_UHRD\_70 (J):  ultra hard band counts at  the source position
on the MOS2 camera within the 70 per cent EEF ellipse.

\noindent 
BKG\_M2\_UHRD\_70 (E):  expected number of ultra  hard band background
counts at  the source position  on the MOS2  camera within the  70 per
cent EEF ellipse.

\noindent 
EXP\_M2\_UHRD\_70 (E):  MOS2 camera ultra  hard band exposure  time in
seconds  at  the  source  position.   This parameter  is  the  average
exposure time within the 70 per cent EEF ellipse.

\noindent 
EML\_DET\_ML\_UHRD  (E): {\sc emldetect}  detection likelihood  in the
ultra hard band for all EPIC cameras.

\noindent 
EML\_EXT\_ML\_UHRD (E): Likelihood that  the source is extended in the
ultra hard  band. This parameter  is estimated by the  {\sc emldetect}
task of SAS for all EPIC cameras.

\noindent 
RA\_CORR (D):  X-ray source Right  Ascension in degrees  corrected for
systematic offsets of the XMM pointing.

\noindent 
DEC\_CORR  (D):  X-ray source  Declination  in  degrees corrected  for
systematic offsets of the XMM pointing.

\noindent 
LII\_CORR   (D):   X-ray   source   Galactic  longitude   in   degrees
corresponding to RA\_CORR and DEC\_CORR

\noindent 
BII\_CORR (D) X-ray source  Galactic latitude in degrees corresponding
to RA\_CORR and DEC\_CORR

\subsection{Source parameters related to flux estimation}


\noindent 
CNT\_TOTAL\_FULL\_80 (J): full  band counts at the source position
from all EPIC cameras estimated within the 80 per cent EEF ellipse.

\noindent 
BKG\_TOTAL\_FULL\_80  (E):  expected number  of  full band  background
counts at the source position from  all EPIC cameras within the 80 per
cent EEF ellipse.

\noindent 
EXP\_TOTAL\_FULL\_80 (E):  full band exposure  time in seconds  at the
source position from  all EPIC cameras. This parameter  is the average
exposure time within the 80 per cent EEF ellipse.

\noindent 
CNT\_PN\_FULL\_80 (J): full band counts  at the source position on the
PN camera within the 80 per cent EEF ellipse.

\noindent 
BKG\_PN\_FULL\_80 (E): expected number  of full band background counts
at the  source position on  the PN camera  within the 80 per  cent EEF
ellipse.

\noindent 
EXP\_PN\_FULL\_80 (E): PN camera full band exposure time in seconds at
the  source position.   This parameter  is the  average  exposure time
within the 80 per cent EEF ellipse.

\noindent 
CNT\_M1\_FULL\_80 (J): full band counts  at the source position on the
MOS1 camera within the 80 per cent EEF ellipse.

\noindent 
BKG\_M1\_FULL\_80 (E): expected number  of full band background counts
at the source  position on the MOS1 camera within the  80 per cent EEF
ellipse.

\noindent 
EXP\_M1\_FULL\_80 (E): MOS1 camera  full band exposure time in seconds
at the source  position.  This parameter is the  average exposure time
within the 80 per cent EEF ellipse.

\noindent 
CNT\_M2\_FULL\_80 (J): full band counts  at the source position on the
MOS2 camera within the 80 per cent EEF ellipse.

\noindent 
BKG\_M2\_FULL\_80 (E): expected number  of full band background counts
at the source  position on the MOS2 camera within the  80 per cent EEF
ellipse.

\noindent 
EXP\_M2\_FULL\_80 (E): MOS2 camera  full band exposure time in seconds
at the source  position.  This parameter is the  average exposure time
within the 80 per cent EEF ellipse.

\noindent 
CNT\_TOTAL\_SOFT\_80 (J): soft  band counts at the source position
from all EPIC cameras estimated within the 80 per cent EEF ellipse.

\noindent 
BKG\_TOTAL\_SOFT\_80  (E):  expected number  of  soft band  background
counts at the source position from  all EPIC cameras within the 80 per
cent EEF ellipse.

\noindent 
EXP\_TOTAL\_SOFT\_80 (E):  soft band exposure  time in seconds  at the
source position from  all EPIC cameras. This parameter  is the average
exposure time within the 80 per cent EEF ellipse.

\noindent 
CNT\_PN\_SOFT\_80  (J): soft band counts  at the source position on the
PN camera within the 80 per cent EEF ellipse.

\noindent 
BKG\_PN\_SOFT\_80  (E): expected number  of soft band background counts
at the  source position on  the PN camera  within the 80 per  cent EEF
ellipse.

\noindent 
EXP\_PN\_SOFT\_80 (E): PN camera soft band exposure time in seconds at
the  source position.   This parameter  is the  average  exposure time
within the 80 per cent EEF ellipse.

\noindent 
CNT\_M1\_SOFT\_80  (J): soft band counts  at the source position on the
MOS1 camera within the 80 per cent EEF ellipse.

\noindent 
BKG\_M1\_SOFT\_80 (E): expected number  of soft band background counts
at the source  position on the MOS1 camera within the  80 per cent EEF
ellipse.

\noindent 
EXP\_M1\_SOFT\_80  (E): MOS1 camera  soft band exposure time in seconds
at the source  position.  This parameter is the  average exposure time
within the 80 per cent EEF ellipse.

\noindent 
CNT\_M2\_SOFT\_80 (J): soft band counts  at the source position on the
MOS2 camera within the 80 per cent EEF ellipse.

\noindent 
BKG\_M2\_SOFT\_80 (E): expected number  of soft band background counts
at the source  position on the MOS2 camera within the  80 per cent EEF
ellipse.

\noindent 
EXP\_M2\_SOFT\_80 (E): MOS2 camera soft band exposure time in seconds
at the source  position.  This parameter is the  average exposure time
within the 80 per cent EEF ellipse.

\noindent 
CNT\_TOTAL\_HARD\_80 (J): hard band counts at the source position
from all EPIC cameras estimated within the 80 per cent EEF ellipse.

\noindent 
BKG\_TOTAL\_HARD\_80  (E):  expected number  of  hard band  background
counts at the source position from  all EPIC cameras within the 80 per
cent EEF ellipse.

\noindent 
EXP\_TOTAL\_HARD\_80 (E):  hard band exposure  time in seconds  at the
source position from  all EPIC cameras. This parameter  is the average
exposure time within the 80 per cent EEF ellipse.

\noindent 
CNT\_PN\_HARD\_80 (J): hard band counts  at the source position on the
PN camera within the 80 per cent EEF ellipse.

\noindent 
BKG\_PN\_HARD\_80 (E): expected number  of hard band background counts
at the  source position on  the PN camera  within the 80 per  cent EEF
ellipse.

\noindent 
EXP\_PN\_HARD\_80 (E): PN camera hard band exposure time in seconds at
the  source position.   This parameter  is the  average  exposure time
within the 80 per cent EEF ellipse.

\noindent 
CNT\_M1\_HARD\_80 (J): hard band counts  at the source position on the
MOS1 camera within the 80 per cent EEF ellipse.

\noindent 
BKG\_M1\_HARD\_80 (E): expected number  of hard band background counts
at the source  position on the MOS1 camera within the  80 per cent EEF
ellipse.

\noindent 
EXP\_M1\_HARD\_80 (E): MOS1 camera  hard band exposure time in seconds
at the source  position.  This parameter is the  average exposure time
within the 80 per cent EEF ellipse.

\noindent 
CNT\_M2\_HARD\_80 (J): hard band counts  at the source position on the
MOS2 camera within the 80 per cent EEF ellipse.

\noindent 
BKG\_M2\_HARD\_80 (E): expected number  of hard band background counts
at the source  position on the MOS2 camera within the  80 per cent EEF
ellipse.

\noindent 
EXP\_M2\_HARD\_80 (E): MOS2 camera  hard band exposure time in seconds
at the source  position.  This parameter is the  average exposure time
within the 80 per cent EEF ellipse.

\noindent 
CNT\_TOTAL\_VHRD\_80 (J): very hard band counts at the source position
from all EPIC cameras estimated within the 80 per cent EEF ellipse.

\noindent 
BKG\_TOTAL\_VHRD\_80 (E): expected number of very hard band background
counts at the source position from  all EPIC cameras within the 80 per
cent EEF ellipse.

\noindent 
EXP\_TOTAL\_VHRD\_80 (E):  very hard band exposure time  in seconds at
the  source position  from all  EPIC  cameras. This  parameter is  the
average exposure time within the 80 per cent EEF ellipse.

\noindent 
CNT\_PN\_VHRD\_80 (J): very hard band counts at the source position on
the PN camera within the 80 per cent EEF ellipse.

\noindent 
BKG\_PN\_VHRD\_80 (E):  expected number  of very hard  band background
counts at the source position on  the PN camera within the 80 per cent
EEF ellipse.

\noindent 
EXP\_PN\_VHRD\_80  (E): PN  camera  very hard  band  exposure time  in
seconds  at  the  source  position.   This parameter  is  the  average
exposure time within the 80 per cent EEF ellipse.

\noindent 
CNT\_M1\_VHRD\_80 (J): very hard band counts at the source position on
the MOS1 camera within the 80 per cent EEF ellipse.

\noindent 
BKG\_M1\_VHRD\_80 (E):  expected number  of very hard  band background
counts at  the source position  on the MOS1  camera within the  80 per
cent EEF ellipse.

\noindent 
EXP\_M1\_VHRD\_80  (E): MOS1 camera  very hard  band exposure  time in
seconds  at  the  source  position.   This parameter  is  the  average
exposure time within the 80 per cent EEF ellipse.

\noindent 
CNT\_M2\_VHRD\_80 (J): very hard band counts at the source position on
the MOS2 camera within the 80 per cent EEF ellipse.

\noindent 
BKG\_M2\_VHRD\_80 (E):  expected number  of very hard  band background
counts at  the source position  on the MOS2  camera within the  80 per
cent EEF ellipse.

\noindent 
EXP\_M2\_VHRD\_80  (E): MOS2 camera  very hard  band exposure  time in
seconds  at  the  source  position.   This parameter  is  the  average
exposure time within the 80 per cent EEF ellipse.

\noindent 
CNT\_TOTAL\_UHRD\_80  (J):  ultra  hard  band  counts  at  the  source
position from  all EPIC cameras estimated  within the 80  per cent EEF
ellipse.

\noindent 
BKG\_TOTAL\_UHRD\_80   (E):  expected  number   of  ultra   hard  band
background counts at the source  position from all EPIC cameras within
the 80 per cent EEF ellipse.

\noindent 
EXP\_TOTAL\_UHRD\_80 (E): ultra hard  band exposure time in seconds at
the  source position  from all  EPIC  cameras. This  parameter is  the
average exposure time within the 80 per cent EEF ellipse.

\noindent 
CNT\_PN\_UHRD\_80 (J):  ultra hard band counts at  the source position
on the PN camera within the 80 per cent EEF ellipse.

\noindent 
BKG\_PN\_UHRD\_80 (E):  expected number of ultra  hard band background
counts at the source position on  the PN camera within the 80 per cent
EEF ellipse.

\noindent 
EXP\_PN\_UHRD\_80  (E): PN  camera ultra  hard band  exposure  time in
seconds  at  the  source  position.   This parameter  is  the  average
exposure time within the 80 per cent EEF ellipse.

\noindent 
CNT\_M1\_UHRD\_80 (J):  ultra hard band counts at  the source position
on the MOS1 camera within the 80 per cent EEF ellipse.

\noindent 
BKG\_M1\_UHRD\_80 (E):  expected number of ultra  hard band background
counts at  the source position  on the MOS1  camera within the  80 per
cent EEF ellipse.

\noindent 
EXP\_M1\_UHRD\_80 (E):  MOS1 camera ultra  hard band exposure  time in
seconds  at  the  source  position.   This parameter  is  the  average
exposure time within the 80 per cent EEF ellipse.

\noindent 
CNT\_M2\_UHRD\_80 (J):  ultra hard band counts at  the source position
on the MOS2 camera within the 80 per cent EEF ellipse.

\noindent 
BKG\_M2\_UHRD\_80 (E):  expected number of ultra  hard band background
counts at  the source position  on the MOS2  camera within the  80 per
cent EEF ellipse.

\noindent 
EXP\_M2\_UHRD\_80 (E):  MOS2 camera ultra  hard band exposure  time in
seconds  at  the  source  position.   This parameter  is  the  average
exposure time within the 80 per cent EEF ellipse.

\noindent 
ECF\_PN\_FULL (E): PN full band counts to flux conversion factor in units of
$\rm 10^{11} counts / (erg \, s^{-1} \,cm^{-2})$.

\noindent 
ECF\_M1\_FULL (E): MOS1 full band counts to flux conversion factor in units of
$\rm 10^{11} counts / (erg \, s^{-1} \,cm^{-2})$.

\noindent 
ECF\_M2\_FULL (E): MOS2 full band counts to flux conversion factor in units of
$\rm 10^{11} counts / (erg \, s^{-1} \,cm^{-2})$.

\noindent 
ECF\_PN\_SOFT (E): PN soft band counts to flux conversion factor in units of
$\rm 10^{11} counts / (erg \, s^{-1} \,cm^{-2})$.

\noindent 
ECF\_M1\_SOFT (E): MOS1 soft band counts to flux conversion factor in units of
$\rm 10^{11} counts / (erg \, s^{-1} \,cm^{-2})$. 

\noindent 
ECF\_M2\_SOFT (E): MOS2 soft band counts to flux conversion factor in units of
$\rm 10^{11} counts / (erg \, s^{-1} \,cm^{-2})$.

\noindent 
ECF\_PN\_HARD  (E): PN hard band counts to flux conversion factor in units of
$\rm 10^{11} counts / (erg \, s^{-1} \,cm^{-2})$.

\noindent 
ECF\_M1\_HARD (E): MOS1 hard band counts to flux conversion factor in units of
$\rm 10^{11} counts / (erg \, s^{-1} \,cm^{-2})$. 

\noindent 
ECF\_M2\_HARD (E): MOS2 hard band counts to flux conversion factor in units of
$\rm 10^{11} counts / (erg \, s^{-1} \,cm^{-2})$.

\noindent 
ECF\_PN\_VHRD (E): PN very hard band counts to flux conversion factor
in units of $\rm 10^{11} counts / (erg \, s^{-1} \,cm^{-2})$.

\noindent 
ECF\_M1\_VHRD (E): MOS1 very hard band counts to flux conversion factor in units of
$\rm 10^{11} counts / (erg \, s^{-1} \,cm^{-2})$. 

\noindent 
ECF\_M2\_VHRD (E): MOS2 very hard band counts to flux conversion factor in units of
$\rm 10^{11} counts / (erg \, s^{-1} \,cm^{-2})$.

\noindent 
ECF\_PN\_UHRD (E): PN ultra hard band counts to flux conversion factor in units of
$\rm 10^{11} counts / (erg \, s^{-1} \,cm^{-2})$.

\noindent 
ECF\_M1\_UHRD  MOS1 ultra hard band counts to flux conversion factor in units of
$\rm 10^{11} counts / (erg \, s^{-1} \,cm^{-2})$. 

\noindent 
ECF\_M2\_UHRD  (E): MOS2 ultra hard band counts to flux conversion factor in units of
$\rm 10^{11} counts / (erg \, s^{-1} \,cm^{-2})$.

\noindent 
HR\_PN  (E):  PN  hardness  ratio.   A  value  of  -9.99  for  HR\_PN,
HR\_PN\_UPPER and HR\_PN\_LOWER (for  the definition of the latter two
parameters see  below) means that  the HR\_PN could not  be estimated,
e.g. the source lies outside the PN FOV.

\noindent 
HR\_PN\_UPPER (E): 68 per cent confidence level upper limit in HR\_PN.

\noindent 
HR\_PN\_LOWER (E): 68 per cent confidence level lower limit in HR\_PN.

\noindent 
HR\_M1  (E):  MOS1 hardness  ratio.   A  value  of -9.99  for  HR\_M1,
HR\_M1\_UPPER  and HR\_M1\_LOWER means  that the  HR\_M1 could  not be
estimated, e.g. the source lies outside the MOS1 FOV.

\noindent 
HR\_M1\_UPPER (E): 68 per cent confidence level upper limit in HR\_M1.

\noindent 
HR\_M1\_LOWER (E): 68 per cent confidence level lower limit in HR\_M1.

\noindent 
HR\_M2  (E):  MOS2 hardness  ratio.   A  value  of -9.99  for  HR\_M2,
HR\_M2\_UPPER  and HR\_M2\_LOWER means  that the  HR\_M2 could  not be
estimated, e.g. the source lies outside the MOS2 FOV.

\noindent 
HR\_M2\_UPPER (E): 68 per cent confidence level upper limit in HR\_M2.

\noindent 
HR\_M2\_LOWER (E): 68 per cent confidence level lower limit in HR\_M2.

\noindent 
FLUX\_FULL (E): 0.5-10\,keV flux in $\rm erg\,s^{-1}\,cm^{-2}$. In the
case of upper limit in the flux this parameter is set to zero.

\noindent 
FLUX\_FULL\_LOWER  (E): 0.5-10\,keV flux  lower limit  uncertainty (68
per cent).  In  the case of upper limit in the  flux this parameter is
set to zero.

\noindent 
FLUX\_FULL\_UPPER (E): 0.5-10\,keV flux upper limit (68 per cent)

\noindent 
FLUX\_SOFT (E): 0.5-2\,keV flux in $\rm erg\,s^{-1}\,cm^{-2}$.

\noindent 
FLUX\_SOFT\_LOWER (E): 0.5-2\,keV flux lower limit (68 per cent)

\noindent 
FLUX\_SOFT\_UPPER (E): 0.5-2\,keV flux upper limit (68 per cent)

\noindent 
FLUX\_HARD (E): 2-10\,keV flux in $\rm erg\,s^{-1}\,cm^{-2}$.

\noindent 
FLUX\_HARD\_LOWER (E): 2-10\,keV flux lower limit (68 per cent)

\noindent 
FLUX\_HARD\_UPPER (E): 2-10\,keV flux upper limit (68 per cent)

\noindent 
FLUX\_VHRD (E): 5-10\,keV flux in $\rm erg\,s^{-1}\,cm^{-2}$.

\noindent 
FLUX\_VHRD\_LOWER (E): 5-10\,keV flux lower limit (68 per cent)

\noindent 
FLUX\_VHRD\_UPPER (E): 5-10\,keV flux upper limit (68 per cent)

\noindent 
FLUX\_UHRD (E): 7.5-12\,keV flux in $\rm erg\,s^{-1}\,cm^{-2}$.

\noindent 
FLUX\_UHRD\_LOWER (E): 7.5-12\,keV flux lower limit (68 per cent)

\noindent 
FLUX\_UHRD\_UPPER (E): 7.5-12\,keV flux upper limit (68 per cent)

\noindent 
LG\_NH\_GAL  (E): Base 10  logarithm of  the Galactic  hydrogen column
density  in the  direction of  the source  in units  of  $\rm cm^{-3}$
estimated from the HI map of \cite{Kalberla2005}.

\noindent 
FILTER\_PN (10A): XMM's PN filter.

\noindent 
FILTER\_MOS1 (10A): XMM's MOS1 filter.

\noindent 
FILTER\_MOS1 (10A): XMM's MOS2 filter.

\subsection{X-ray optical identification parameters}

\noindent 
SDSS\_ID (20A): SDSS source identification number.

\noindent 
LR\_SOFT (E): Likelihood of  the optical counterpart for X-ray sources
in the soft-band selected sample.

\noindent 
LR\_HARD (E): Likelihood of  the optical counterpart for X-ray sources
in the hard-band selected sample.

\noindent 
LR\_VHRD (E): Likelihood of  the optical counterpart for X-ray sources
in the very hard band selected sample.

\noindent 
LR\_UHRD (E): Likelihood of  the optical counterpart for X-ray sources
in the ultra hard band selected sample.  sample.

\noindent 
LR\_FULL (E): Likelihood of  the optical counterpart for X-ray sources
in the full band selected sample.

\noindent 
LR  (E): Likelihood  of  the optical  counterpart.  This parameter  is
estimated for all X-ray sources independent of the detection band.

\noindent 
REL\_SOFT  (E):  Reliability  of  the optical  counterpart  for  X-ray
sources in the soft band selected sample.

\noindent 
REL\_HARD  (E):  Reliability  of  the optical  counterpart  for  X-ray
sources in the hard band selected sample.

\noindent 
REL\_VHRD  (E):  Reliability  of  the optical  counterpart  for  X-ray
sources in the very hard band selected sample.

\noindent 
REL\_UHRD  (E):  Reliability  of  the optical  counterpart  for  X-ray
sources in the ultra hard -band selected sample.

\noindent 
REL\_FULL  (E):  Reliability  of  the optical  counterpart  for  X-ray
sources in the full band selected sample.

\noindent 
REL  (E): Reliability of  the optical  counterpart. This  parameter is
estimated for all X-ray sources independent of the detection band.

\noindent 
DR\_ARCSEC (E): angular separation  in arcsec between the X-ray source
and the optical counterpart.

\noindent 
DR\_RA\_ARCSEC (E): RA  offset in arcsec between the  X-ray source and
the optical counterpart.

\noindent 
DR\_DEC\_ARCSEC (E): DEC offset in arcsec between the X-ray source and
the optical counterpart

\subsection{SDSS photometry parameters} 

\noindent 
petro\_u (E): SDSS $u$-band petrosian magnitude of the optical
counterpart.

\noindent 
petro\_g (E): SDSS $g$-band petrosian magnitude of the optical
counterpart.

\noindent 
petro\_r (E): SDSS $r$-band petrosian magnitude of the optical
counterpart.

\noindent 
petro\_i (E): SDSS $i$-band petrosian magnitude of the optical
counterpart.

\noindent 
petro\_z (E): SDSS $z$-band petrosian magnitude of the optical
counterpart.

\noindent 
petro\_err\_u (E): SDSS $u$-band petrosian magnitude error of the
optical counterpart.

\noindent 
petro\_err\_g (E): SDSS $g$-band petrosian magnitude error of the
optical counterpart. 

\noindent 
petro\_err\_r (E): SDSS $r$-band petrosian magnitude error of the
optical counterpart.

\noindent 
petro\_err\_i (E): SDSS $i$-band petrosian magnitude error of the
optical counterpart.

\noindent 
petro\_err\_z (E): SDSS $z$-band petrosian magnitude error of the
optical counterpart.

\noindent 
model\_u (E): SDSS $u$-band model magnitude of the optical
counterpart.

\noindent 
model\_g (E): SDSS $g$-band model magnitude of the optical
counterpart.

\noindent 
model\_r (E): SDSS $r$-band model magnitude of the optical
counterpart.

\noindent 
model\_i (E): SDSS $i$-band model magnitude of the optical
counterpart.

\noindent 
model\_z (E): SDSS $z$-band model magnitude of the optical
counterpart.

\noindent 
model\_err\_u (E): SDSS $u$-band model magnitude error of the
optical counterpart.

\noindent 
model\_err\_g (E): SDSS $g$-band model magnitude error of the
optical counterpart. 

\noindent 
model\_err\_r (E): SDSS $r$-band model magnitude error of the
optical counterpart.

\noindent 
model\_err\_i (E): SDSS $i$-band model magnitude error of the
optical counterpart.

\noindent 
model\_err\_z (E): SDSS $z$-band model magnitude error of the
optical counterpart.

\noindent 
psf\_u (E): SDSS $u$-band psf magnitude of the optical
counterpart 

\noindent 
psf\_g (E): SDSS $g$-band psf magnitude of the optical
counterpart

\noindent 
psf\_r (E): SDSS $r$-band psf magnitude of the optical
counterpart

\noindent 
psf\_i (E): SDSS $i$-band psf magnitude of the optical
counterpart

\noindent 
psf\_z (E): SDSS $z$-band psf magnitude of the optical
counterpart.

\noindent 
psf\_err\_u (E): SDSS $u$-band psf magnitude error of the
optical counterpart. 

\noindent 
psf\_err\_g (E): SDSS $u$-band psf magnitude error of the
optical counterpart.

\noindent 
psf\_err\_r (E): SDSS $u$-band psf magnitude error of the
optical counterpart.

\noindent 
psf\_err\_i (E): SDSS $u$-band psf magnitude error of the
optical counterpart. 

\noindent 
psf\_err\_z (E): SDSS $u$-band psf magnitude error of the
optical counterpart.

\noindent 
ext\_u (E): Galactic extinction in magnitudes in the SDSS $u$-band.  

\noindent 
ext\_g (E): Galactic extinction in magnitudes in the SDSS $g$-band.

\noindent 
ext\_r (E): Galactic extinction in magnitudes in the SDSS $r$-band.

\noindent 
ext\_i (E): Galactic extinction in magnitudes in the SDSS $i$-band.

\noindent 
ext\_z (E): Galactic extinction in magnitudes in the SDSS $z$-band.

\noindent 
p\_type (J): SDSS photometric type, 3 for optically extended source, 6
for stellar object.     

\noindent 
GAL\_LAT (D): Galactic latitude of the counterpart.

\noindent 
GAL\_LONG (D): Galactic longitude of the counterpart.

\subsection{SDSS photometric and spectroscopic redshift  parameters} 

\noindent 
SDSS\_SPEC\_ID (20A): SDSS spectroscopic identification number, if
available. 

\noindent 
z (D): spectroscopic redshift

\noindent 
ztype(D): spectroscopic redshift type

\noindent 
z\_phot1 (D): Neural Netwotk photometric redshift estimated by the 
CC2 method described by \cite{Oyaizu2008}.

\noindent 
z\_err\_1 (D): photometric redshift error for the CC2 method of  \cite{Oyaizu2008}. 

\noindent 
z\_phot2 (D): Neural Netwotk photometric redshift estimated by the 
D1 method described by \cite{Oyaizu2008}.

\noindent 
z\_err\_2  (D): photometric redshift error for the D1 method of  \cite{Oyaizu2008}.


\section{Simulations: the CMD of X-ray sources in the AEGIS as would
  appear at low redshift}\label{sec_simulations} 

Differential selection effects  in the comparison of the  CMD of X-ray
AGN  in  the AEGIS sample and  the low redshift subset of the XMM/SDSS
survey  
are  accounted for  by
performing simulations  to explore how  the rest-frame colours  of the
AEGIS X-ray sources at $z\approx0.8$  would appear in our low redshift
($0.03<z<0.2$)  XMM/SDSS  survey.  We  choose  to  transform the  high
redshift  sample  to  low  redshifts.   This is  because  the  Chandra
observations  of the  AEGIS field  select AGN  at  rest-frame energies
1-14\,keV at $z\approx0.8$.  That sample is therefore less affected by
obscuration compared  to the XMM/SDSS survey, which  probes lower rest
frame energies ($\approx 2-8$\,keV) at $z<0.2$. The simulations proceed
as  follows. AEGIS X-ray  sources are  randomly selected  by weighting
each one  by $1/V_{max}$, the  inverse of the maximum  comoving volume
within which  it can be detected.   Each source is placed  at a random
redshift in  the range $0.03-0.2$  and its SDSS $r$-band  magnitude is
estimated using  the {\sc kcorrect} routines.  The  source is retained
in the sample if it is brighter than the magnitude limit $r=17.77$\,mag
of  the  XMM/SDSS low  redshift  survey.   The  2-10\,keV X-ray  flux,
$f_X(\rm 2-10\,keV)$, of the source  is also determined from the X-ray
luminosity,  $L_X(\rm  0.5-10\,keV)$,  and  hydrogen  column  density,
$N_H$.  The sensitivity curve of the XMM/SDSS survey is an estimate of
the  probability of  detecting a  source  with given  flux within  the
surveyed area.  A random number  is produced between 0 and the maximum
value of  the XMM/SDSS survey 2-10\,keV sensitivity  curve.  If the
random number is lower than the value of the sensitivity 
curve  at flux  $f_X(\rm 2-10\,keV)$  the  source is  retained in  the
sample.  A total of 10,000 realisations are performed.

The estimation  of the maximum  volume takes into account  the optical
magnitude  limits for spectroscopy  in the  AEGIS, $18.5<R<24.1$\,mag,
the  varying X-ray  sensitivity in  that field  and  the spectroscopic
sampling rate.  For  source $i$ at redshift $z$,  with intrinsic X-ray
luminosity $L_X$, obscuring hydrogen column density $N_H$ and absolute
optical magnitude $M_{^{0.1}}$ the maximum volume is

\begin{equation} V_{max,i}(L_X,N_H,M,z) = \frac{c}{H_0} \int_{z1}^{z2} \,
\Omega(L_X,z)\, \, w_{i} \,\frac{dV}{dz}\,dz\, dL,
\end{equation}

\noindent where  $dV/dz$ is the  volume element per  redshift interval
$dz$ and $w_{i}$ is a weight that depends on the observed $B-R$, $R-I$
colours and the  $R$-band magnitude of the X-ray  sources and corrects
for  the spectroscopic  sampling rate  of the  X-ray  population.  The
weights are estimated using the ``minimal'' model described by \cite{Willmer2006}. The  integration limits are  $z1=max(z_{bright}, z_L)$
and $z2=min(z_{faint},  z_U)$, where we have defined  $z_L$, $z_U$ the
lower  and  upper  redshift  limits  of the  sample  and  $z_{faint}$,
$z_{bright}$  the redshifts at  which the  source will  become fainter
than  $R=24.1$\,mag  and  brighter  than  $R=18.5$\,mag,  the  optical
magnitude  limits  for optical  spectroscopy.  $\Omega(L_X,z)$ is  the
solid angle of the X-ray  survey available to a source with luminosity
$L_X$ and  column density  $N_H$ at redshift  $z$ (corresponding  to a
flux $f_X$  in the  X-ray sensitivity curve).   The estimation  of the
X-ray flux from $L_X$ takes into account the intrinsic source $N_H$ by
adopting a  power-law spectral energy  distribution with $\Gamma=1.9$
and photoelectric    absorption   cross    sections   as    described   by
\cite{Morrison1983} for solar metallicity.

\end{document}